 \documentclass[]{elsart5p}

\usepackage{natbib}

 \usepackage{graphicx}
\usepackage{amssymb} \usepackage{rotating}

\def\astrobj#1{#1}

\def\l{$\lambda$}
\def\ll{$\lambda\lambda$}

\def\hea{He\,{\sc i}}
\def\heb{He\,{\sc ii}}
\def\caa{Ca\,{\sc i}}
\def\cab{Ca\,{\sc ii}}

\def\nc{N\,{\sc iii}}
\def\naa{Na\,{\sc i}}

\def\oc{O\,{\sc iii}}

\def\cc{C\,{\sc iii}}
\def\cd{C\,{\sc iv}}
\def\sic{Si\,{\sc iii}}
\def\sid{Si\,{\sc iv}}

\def\sun{\odot}
\def\kms{km\,s$^{-1}$}
\def\cnts{cnt\,s$^{-1}$}
\def\ergs{erg\,s$^{-1}$}

\def\rsol{R$_{\sun}$}
\def\msol{M$_{\sun}$}
\def\msolyr{M$_{\sun}$~yr$^{-1}$}
\def\degr{$^\circ$}

\newcommand\fd{\hbox{$.\!\!^{\reset@font\romn d}$}}
\newcommand\fh{\hbox{$.\!\!^{\reset@font\romn h}$}}
\newcommand\fm{\hbox{$.\!\!^{\reset@font\romn m}$}}
\newcommand\fs{\hbox{$.\!\!^{\reset@font\romn s}$}}
\newcommand\fdg{\hbox{$.\!\!^\circ$}}
\newcommand\farcm{\hbox{$.\mkern-4mu^\prime$}}
\newcommand\farcs{\hbox{$.\!\!^{\prime\prime}$}}
\newcommand\fp{\hbox{$.\!\!^{\reset@font\reset@font\scriptscriptstyle\romn p}$}}

\def\xmm{{\sc XMM}\emph{-Newton}}

\def\epic{{\sc EPIC}}
\def\mos{{\sc MOS}}
\def\pn{pn}

\def\epicmos{{\sc EPIC MOS}}

\def\fer{{\sc FEROS}}

\def\ces{{\sc CES}}
\def\bme{{\sc BME}}
\def\mid{{\sc MIDAS}}

\def\aap{A\&A}
\def\aaps{A\&AS}
\def\aj{AJ}
\def\apj{ApJ}
\def\apjs{ApJS}

\def\aplett{Astrophys. Lett.}

\def\mnras{MNRAS}

\def\pasp{PASP}

\def\xspec{{\sc xspec}}
\def\mek{{\tt mekal}}

\def\eml{{\it emldetect}}
\def\arf{{\it arf}}
\def\rmf{{\it rmf}}

\defcitealias{HCB74}{HCB74}
\defcitealias{LM83}{LM83}
\defcitealias{HMM85}{HMM85}
\defcitealias{PHYB90}{PHYB90}
\defcitealias{GM01}{GM01}
\defcitealias{SLP97}{SLP97}

\begin{document}

\begin{frontmatter}



\title{The massive binary HD~152218 revisited\thanksref{titlenote}: a new colliding wind system in NGC~6231}
\thanks[titlenote]{Based on observations collected at the European Southern Observatory (La Silla, Chile), at the Cerro Tololo Inter-American Observatory (Cerro Tololo, Chile) and with \xmm, an ESA Science Mission with instruments and contributions  directly funded by ESA Member States and the USA (NASA).}


\author[ESO]{H. Sana}
\author[Liege]{Y. Naz\'e\corauthref{FNRS}}
\author[UK,Liege]{B. O'Donnell}
\author[Liege]{G. Rauw\corauthref{FNRS}\corauthref{ctio}}
\author[Liege]{E. Gosset\corauthref{FNRS}\corauthref{ctio}}

\corauth[FNRS]{FNRS, Belgium}
\corauth[ctio]{Visiting Astronomer, CTIO, National Optical Astronomy Observatories (NOAO). NOAO is operated by the Association of Universities for Research in Astronomy, Inc.\ under contract with the National Science Foundation.}

\address[ESO]{European Southern Observatory, Alonso de Cordova 3107, Vitacura, Santiago 19, Chile} 
\address[Liege]{Institut d'Astrophysique et de G\'eophysique, Li\`ege University, All\'ee du 6 Ao\^ut 17, Bat. B5c, B-4000 Li\`ege, Belgium }
\address[UK]{Dept. of Physics \& Astronomy, University College London, Gower Street, London WC1E 6BT, UK.}

\begin{abstract}
We present the results of an optical and X-ray  monitoring campaign on the short-period massive SB2 binary HD~152218.  Combining our HiRes spectroscopic data with previous observations, we unveil the contradictions between the published orbital solutions. In particular, we solve the aliasing on the period and derive a value close to 5.604~d. Our eccentricity $e=0.259\pm0.006$ is slightly lower than previously admitted. We show that HD~152218 is probably undergoing a relatively rapid apsidal motion of about 3\degr~yr$^{-1}$ and we confirm the O9IV+O9.7V classification. We derive minimal masses of $15.82 \pm 0.26$~\msol\ and   $12.00 \pm 0.19$~\msol\ and constrain the radius of the components to  $R_1=10.3\pm1.3$~\rsol\ and $R_2=7.8\pm1.7$~\rsol. 
We also report the results of a \xmm\  monitoring of the  HD~152218 X-ray emission throughout its orbital motion. The averaged X-ray spectrum is relatively soft and it is well reproduced by a 2-T optically thin thermal plasma model with component temperatures about 0.3 and 0.7~keV. The system presents an increase of its  X-ray flux by about 30\% near apastron compared to periastron, which is interpreted as the signature of an ongoing wind-wind interaction process occurring within the wind acceleration region. 
\end{abstract}


\end{frontmatter}

% main text

\begin{table*}
\begin{center}
\caption{
Journal of the spectroscopic observations of HD~152218. Column 1  lists the Heliocentric Julian Date (in format HJD$-$2\,450\,000) at mid-exposure. The next three columns present the phases (calculated from the \hea-line SB2 orbital solution of Table \ref{tab: orbit}) and the averaged primary and secondary radial velocities computed in the respective systemic velocity frames (see Sect. \ref{ssect: orb}). Instrumental setups are given at the bottom of the table.}
\label{tab: opt_diary} 
\begin{tabular}{r r r r | r r r r}
\hline
HJD & $\phi_\mathrm{HeI}$  & $\overline{RV_{\lambda,1}-\gamma_{\lambda,1}}$ &  $\overline{RV_{\lambda,2}-\gamma_{\lambda,2}}$ & HJD & $\phi_\mathrm{HeI}$  & $\overline{RV_{\lambda,1}-\gamma_{\lambda,1}}$ &  $\overline{RV_{\lambda,2}-\gamma_{\lambda,2}}$  \\
     (days)    &           &  (\kms)   &  (\kms)   &  (days)        &           &  (\kms)   &  (\kms)     \\
\hline
 621.673$^a$   &   0.406   & $-$25.00  &  $-$29.31 & 1302.901$^c$   &   0.968   &  $-$2.09  &      0.46    \\
 622.642$^a$   &   0.579   &    63.69  & $-$133.73 & 1304.902$^c$   &   0.325   & $-$82.39  &    134.85    \\
 622.755$^a$   &   0.599   &    80.30  & $-$123.20 & 1323.853$^c$   &   0.707   &   132.51  & $-$175.86    \\
 623.648$^a$   &   0.759   &   151.20  & $-$200.79 & 1327.796$^c$   &   0.411   & $-$11.80  &   $-$9.25    \\
 623.762$^a$   &   0.779   &   163.30  & $-$188.40 & 1328.905$^b$   &   0.609   &    82.27  & $-$124.90    \\
 624.590$^a$   &   0.927   &    74.35  & $-$116.83 & 1329.855$^b$   &   0.778   &   147.17  & $-$199.78    \\
 624.722$^a$   &   0.950   &    13.36  &      9.05 & 1331.882$^b$   &   0.140   &$-$182.39  &    213.02    \\
 625.626$^a$   &   0.112   &$-$158.93  &    209.21 & 1669.812$^c$   &   0.442   &  $-$6.42  &   $-$3.87    \\
 625.734$^a$   &   0.131   &$-$171.70  &    233.10 & 1670.804$^c$   &   0.619   &    96.65  & $-$114.87    \\
 626.591$^a$   &   0.284   &$-$104.57  &    166.73 & 1670.918$^c$   &   0.639   &   108.37  & $-$132.91    \\
 626.716$^a$   &   0.306   & $-$85.00  &    141.40 & 1671.828$^c$   &   0.802   &   153.10  & $-$200.44    \\
1299.883$^c$   &   0.430   &  $-$7.56  &   $-$5.01 & 1672.865$^c$   &   0.987   & $-$12.61  &  $-$10.06    \\
1300.882$^c$   &   0.608   &    94.15  & $-$107.22 & 1673.878$^c$   &   0.168   &$-$168.93  &    211.74    \\
1301.883$^c$   &   0.787   &   152.44  & $-$201.56 & 2037.812$^c$   &   0.110   &$-$172.15  &    221.33    \\
\hline
\end{tabular}
\end{center}
\end{table*}

\section{Introduction} \label{sect: intro}

Early-type stars of spectral type O are characterized by strong stellar winds. Though not as extreme as those of  their evolved descendants, the Wolf-Rayet stars, these winds combine  terminal velocities of a few thousand \kms\ and important mass-loss rates (about $10^{-7}-10^{-5}$~\msol~yr$^{-1}$) that significantly affect both the surroundings of the star and its evolution. In a binary system, it is expected that the winds from the two stars collide, producing a density enhanced region known as the wind interaction zone. The shock-heated plasma within this zone is expected to produce an additional contribution to the X-ray emission from the early-type system and, indeed, the early-type binaries are known to be statistically X-ray overluminous compared to single stars of the same spectral type \citep{ChG91}. This extra X-ray emission might display phase-locked modulations, either due to a change of the absorption properties along the line of sight, or to a modulation of the shock strength due e.g. to a varying separation between the components in an eccentric binary. The properties of the shock thus strongly depend both on the geometry of the binary system and on the characteristics of the individual winds. However these properties often cruelly lack accurate constraints. In this framework, our team is involved in a long-standing effort to study early-type stars in a number of open clusters. 

The present work focuses on the  short-period binary \astrobj{HD152218}, located in \astrobj{NGC 6231} at the core of the \astrobj{Sco OB 1} association. This star is a long known SB2 system \citep{Str44}. Previous papers \citep{HCB74, LM83, SLP97, GM01} however reported conflicting results and we thus decided to re-appraise the orbital properties of this apparently well-studied system. The aim is to bring  its orbital and physical parameters on firm ground. As mentioned above, accurate ephemeris and a detailed knowledge of the orbital and physical properties of a system are indeed crucial ingredients for interpreting the X-ray data and for uncovering the possible signature of a wind interaction.

The remaining of this paper is organised as follows. The next section describes the observational material and the data reduction processes. Sect. \ref{sect: orbit} aims at unveiling the discrepancies of the previously published orbital solutions. Sect. \ref{sect: physic} discusses the  physical properties and evolutionary status of HD~152218 and, in Sect. \ref{sect: xmm}, we analyse the X-ray observations in the light of the upgraded orbital solution. Finally Sect. \ref{sect: ccl} provides a summary of the present work.

%_____________________________ DATA REDUCTION ______________________________
\section{Observations and data reduction }\label{sect: obs} 

\subsection{Optical spectroscopy }\label{ssect: obs_os}

The present work is based on 42 high-resolution spectra of \astrobj{HD 152218} obtained at the European Southern Observatory (ESO, La Silla, Chile) and at the Cerro Tololo Inter-American Observatory (CTIO). These were acquired during 35 different nights distributed over 8 runs between May 1997 and May 2004. The journal of the observations is presented in Table \ref{tab: opt_diary}.		
					     
In May 1997, five (resp. six) high resolution spectra of the \hea\,\l4471 (resp. \heb\,\l4686) line were obtained with ESO's 1.4\,m Coud\'e Auxiliary Telescope (CAT) at La Silla, using the Coud\'e Echelle Spectrometer (CES) equipped with the Long Camera (LC). The detector used was ESO CCD\#38, a Loral $2688 \times 512$ pixel CCD with a pixel size of $15\times 15\,\mu$m. The slit width was chosen to achieve a nominal resolving power of 70000--80000. The effective resolving power as derived from the {\sc FWHM} of the lines of the ThAr calibration exposures is 65000--75000. Typical exposure times were of the order of 30 minutes and the average signal-to-noise ratio (SNR) is about 140. The observed wavelength domain is centered on the \hea\,\l4471 or \heb\,\l4686 line and is $\sim$ 45\,\AA\ wide.

Another set of three echelle spectra over the range 3850 to 5790\,\AA\ was obtained with the Bench-Mounted Echelle Spectrograph (BME) attached to the 1.5\,m CTIO Ritchey-Chr\'etien Telescope, during a 5-night run in June 1999. Forty nine orders were observed using the KPGL2 316\,lines\,mm$^{-1}$ grating as a cross-disperser. The detector was a Tek\,2048 CCD with 24\,$\mu$m pixels. The slit width was set to 70\,$\mu$m corresponding to a resolving power of 45000. Exposure times ranged from 45 to 60~min, yielding a SNR in the continuum between 40 and 50.\\

Between April 1999 and May 2002, we collected 20 echelle spectra covering the whole optical range ($\sim$3750--9200\AA) using the Fiber-fed Extended Range Optical Spectrograph (\fer), an echelle spectrograph mounted at the ESO\,1.5m telescope at La Silla. In May 2003 and May 2004, respectively three and five other FEROS spectra were obtained at the ESO\,2.2m telescope, at La Silla too. The detector was a 2k $\times$ 4k EEV CCD with a pixel size of 15$\mu$m $\times$ 15$\mu$m. The spectral resolving power of \fer\ is 48\,000. Typical exposure times ranged, at the ESO\,1.5m telescope, from 10 to 20\,min according to the weather conditions, resulting in typical SNR around 150. At the ESO\,2.2m telescope, the SNR is about 200 in May 2003 and about 250 in May 2004. Typical exposure times at the ESO\,2.2m were of 13 min. \\

%%%%%%%%%%%%%%%%%%%%%%%%%%%%%%%%%%%%%%%%%%%%%%%%%%%%%%%%%%%%%%%%%%%%%%%%%
\begin{table}
\begin{center}
 Table 1. {\it Continue} \\
\begin{tabular}{r r r r }
\hline
HJD & $\phi_\mathrm{HeI}$  & $\overline{RV_{\lambda,1}-\gamma_{\lambda,1}}$ &  $\overline{RV_{\lambda,2}-\gamma_{\lambda,2}}$ \\

   (days)     &          &  (\kms)   & (\kms)      \\ 
\hline
 2037.903$^c$ &   0.126   &$-$174.37  &    222.99   \\
 2039.800$^c$ &   0.465   &  $-$3.87  &   $-$1.32   \\
 2040.859$^c$ &   0.654   &   111.51  & $-$142.01   \\
 2381.695$^c$ &   0.474   &  $-$2.42  &      0.13   \\
 2382.692$^c$ &   0.652   &   109.41  & $-$147.06   \\
 2383.691$^c$ &   0.831   &   149.50  & $-$193.52   \\
 2782.709$^d$ &   0.033   & $-$98.31  &    129.27   \\
 2783.748$^d$ &   0.219   &$-$148.99  &    194.69   \\
 2784.699$^d$ &   0.389   & $-$17.66  &  $-$15.11   \\
 3130.843$^d$ &   0.156   &$-$173.97  &    223.41   \\
 3131.716$^d$ &   0.312   & $-$85.25  &    133.87   \\
 3132.730$^d$ &   0.493   &     4.66  &      3.89   \\
 3133.760$^d$ &   0.677   &   115.83  & $-$163.34   \\
 3134.723$^d$ &   0.849   &   141.10  & $-$189.24   \\
\hline
\end{tabular}\\ 
\begin{flushleft}
$^a$ CES + ESO-CAT; $^b$ BME + CTIO~1.5m ; $^c$ FEROS + ESO~1.5m; $^d$ FEROS + ESO/MPG~2.2m
\end{flushleft}
\end{center}
\end{table}
%%%%%%%%%%%%%%%%%%%%%%%%%%%%%%%%%%%%%%%%%%%%%%%%%%%%%%%%%%%%%%%%%%%%%%%%%

%%%%%%%%%%%%%%%%%%%%%%%%%%%%%%%%%%%%%%%%%%%%%%%%%%%%%%%%%%%%%%%%%%%%%%%%%
\begin{table*}
\caption{Journal of the \xmm\ observations of \astrobj{HD 152218}. The  Julian Date (JD) at mid-exposure is given in Col. 2. Cols. 3 and 4  list the effective exposure times for the  \epicmos1 and \mos2 instruments while Cols. 5 and 6 report the background-subtracted, vignetting and exposure-corrected count-rates in the different instruments. The last column provides the orbital phase of \astrobj{HD 152218} for each \xmm\ observation at mid-exposure, according to the \hea-line ephemerides given in Table \ref{tab: orbit}. The phase extension of the pointings is about 0.06. Obs. 2 and 4 were however affected by high background events that reduced the nominal exposure time by about one third.
}
\label{tab: journal} 
\centering          

\begin{tabular}{c c c c  c c c}
\hline
\vspace*{-3mm}\\
Obs. \# & JD &  \multicolumn{2}{c}{Effective duration (ks)} &  \multicolumn{2}{c}{Count rates ($10^{-3}$ \cnts)} & Phase $\phi_\mathrm{HeI}$ of\\
       &JD$-2\,450\,000$& \mos1 & \mos2   & \mos1 & \mos2  & \astrobj{HD 152218} \\ 
\hline
1 &  2158.214 & 33.1 & 33.2 & $13.3 \pm 0.8 $ & $14.7 \pm 0.9 $ &  0.591  \\
2 &  2158.931 & 19.8 & 19.8 & $17.5 \pm 1.2 $ & $12.9 \pm 1.0 $ &  0.719  \\
3 &  2159.796 & 33.7 & 33.9 & $13.8 \pm 0.8 $ & $13.1 \pm 0.8 $ &  0.873  \\
4 &  2160.925 & 26.0 & 24.3 & $12.5 \pm 0.9 $ & $11.6 \pm 0.9 $ &  0.075  \\
5 &  2161.774 & 30.9 & 31.0 & $14.4 \pm 0.9 $ & $13.4 \pm 0.9 $ &  0.226  \\
6 &  2162.726 & 32.9 & 32.8 & $15.7 \pm 0.9 $ & $14.4 \pm 0.9 $ &  0.396  \\
\hline
\end{tabular}
\end{table*}
%%%%%%%%%%%%%%%%%%%%%%%%%%%%%%%%%%%%%%%%%%%%%%%%%%%%%%%%%%%%%%%%%%%%%%%%%

The \ces\ data were reduced in a standard way using the \mid\ package supported by ESO. The spectra were rectified by means of an instrumental response curve built from the observations, obtained under similar conditions, of a metal-poor `reference' star (HD 203608: F6V). Finally, the spectra were normalized by fitting a low order polynomial  to the continuum. 

The BME data were reduced using the IRAF\footnote{IRAF is distributed by the National Optical Astronomy Observatories.} package and following the recommendations of the BME User's Manual. The pixel to pixel variations were removed using flat field exposures taken with a very bright light source and a diffusing screen placed inside the spectrograph (so-called {\it milky flats}). A first rectification of the extracted echelle orders was carried out with the projector flat exposures. The spectra were then normalized by fitting a low-order polynomial to the continuum. 

\fer\ data were reduced using an improved version of the \fer\ context \citep[see details in ][]{SHRG03} working under the \mid\ environment. We mainly used the normalized individual orders. However, when doubts about the quality of the normalisation arise, we checked our results with the automatically merged spectrum.

%_____________________ X-ray OBSERVATIONS ________________________________

\subsection{X-ray observations } \label{ssect: xmm_obs}

In September 2001, the \astrobj{NGC 6231} open cluster was the target of an X-ray observing campaign with the \xmm\ European observatory \citep{jansen01_xmm}. The campaign, of a total duration of about 180\,ks, is described in length in \citet{SGR06} together with the data reduction and source identification. We only give a brief overview here, focusing on the additional elements specific to \astrobj{HD 152218}. 

\begin{figure}
   \centering
   \includegraphics[width=.9\columnwidth]{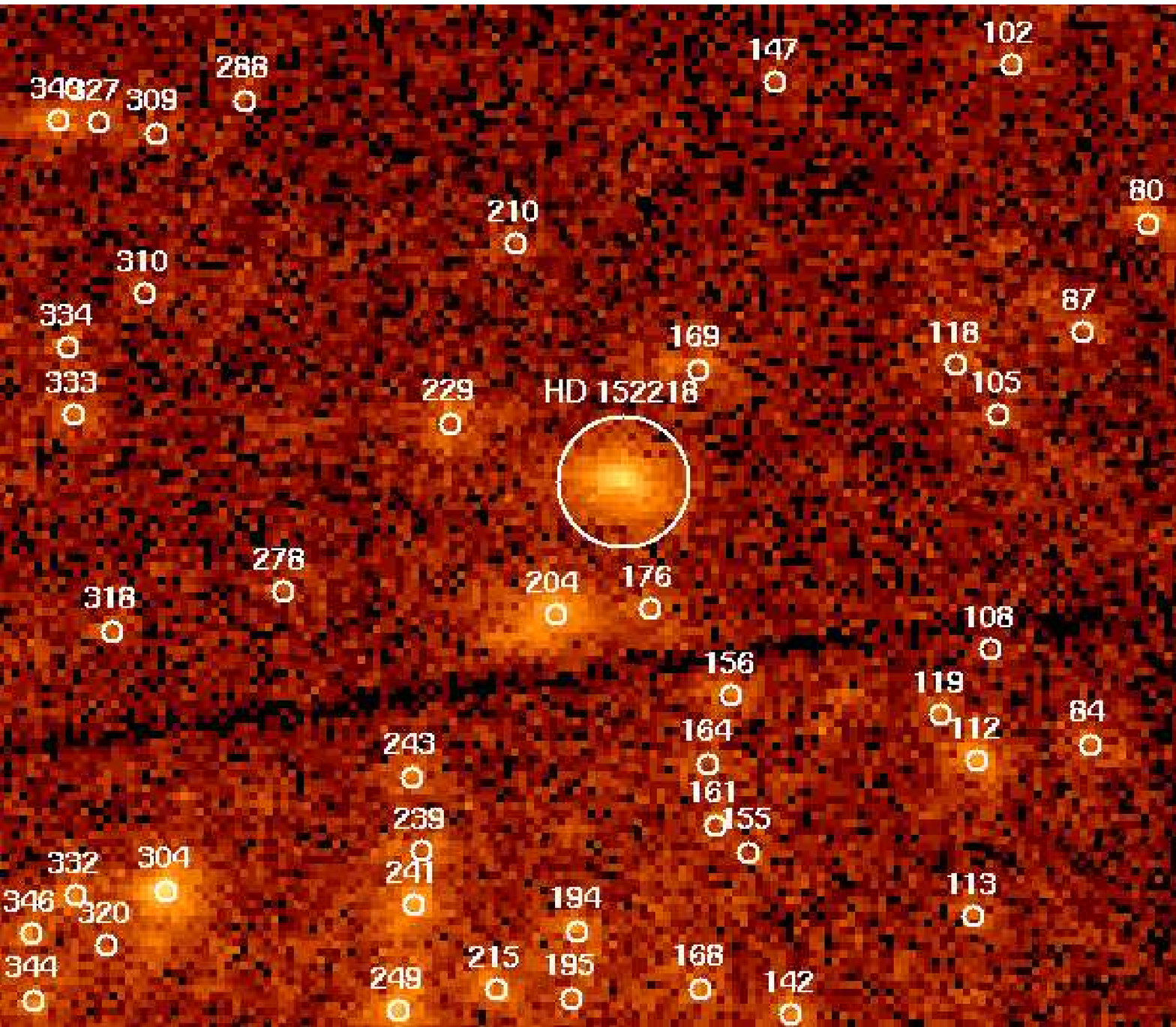}
   \caption{Combined \epicmos1 + \mos2 image in the vicinity of \astrobj{HD 152218}. The adopted source extraction region is shown. Neighbouring X-ray sources are labelled using the internal numbering of \citet{SGR06}.} 
   \label{fig: mos} 
\end{figure}

The campaign actually consisted of six separate observations spread over 5 days.  All three \epic\ instruments \citep{Struder01_pn_mnras,Turner01_mos_mnras} were operated in the Full Frame mode together with the Thick Filter to reject UV/optical light. The field of view (FOV) of the \epic\ instruments was centered on the colliding wind binary HD~152248, from which \astrobj{HD 152218} is separated by about 6\farcm8. 
 Due to the brightness of the cluster objects in the FOV, the Optical Monitor \citep{Mason01_om} was switched off throughout the campaign. 
Fig. \ref{fig: mos} presents a view of the combined \mos\ images around \astrobj{HD 152218} while Table \ref{tab: journal} provides the journal of the X-ray observations of the system.

\astrobj{HD 152218} unfortunately fell on a gap of the \pn\ instrument and, as in \citet{SGR06}, we only considered the \epicmos\ data. To obtain the averaged count-rates during the different pointings, we adopted the source positions given in the catalogue of \citet{SGR06} and we performed a psf (point spread function) fit using the SAS task \eml. As in the previous works, we restrained our scientific analysis to the 0.5-10.0~keV band and we adopted three energy sub-ranges: a soft (S$_\mathrm{X}$) [0.5:1.0~keV], a medium (M$_\mathrm{X}$) [1.0:2.5~keV] and a hard (H$_\mathrm{X}$) [2.5-10.0~keV] band. 

Located slightly offset from the cluster core, \astrobj{HD 152218} is reasonably well isolated in the X-ray images. To extract the source spectra, we used an extraction radius of 20\farcs2, centered on the source position for all six pointings and for the two \epicmos\ instruments.  The different backgrounds were estimated from  source-free regions  located on the same CCD detector as \astrobj{HD 152218} (CCD \#3 and 5 respectively for the \mos1 and \mos2 instruments).  The adopted source extraction region is shown in Fig. \ref{fig: mos}. Displays of the corresponding background extraction regions are presented in \citet{SRN06} together with their positions and sizes.

Using these regions, we extracted the source spectra from the two \mos\ instruments and for the six pointings individually. We also extracted the merged spectra in each instrument, thus combining the six observations of \astrobj{HD 152218}. For this purpose, we built the corresponding Ancillary Response Files (the so-called \arf\ files) using the SAS task {\it arfgen}. We adopted the Redistribution Matrix Files (\rmf) provided by the SOC. The spectra were finally binned to reach at least 10 counts per bin. The X-ray data will be discussed in Sect. \ref{sect: xmm}.

% ************************************************************************* 
% *************************************************************************

%_______________________ ORBITAL SOLUTION __________________________

\section{Unveiling the correct orbital solution} \label{sect: orbit}

\subsection{Optical spectrum} \label{ssect: RVs}

\begin{figure*}
\begin{center}
\includegraphics[width=\textwidth]{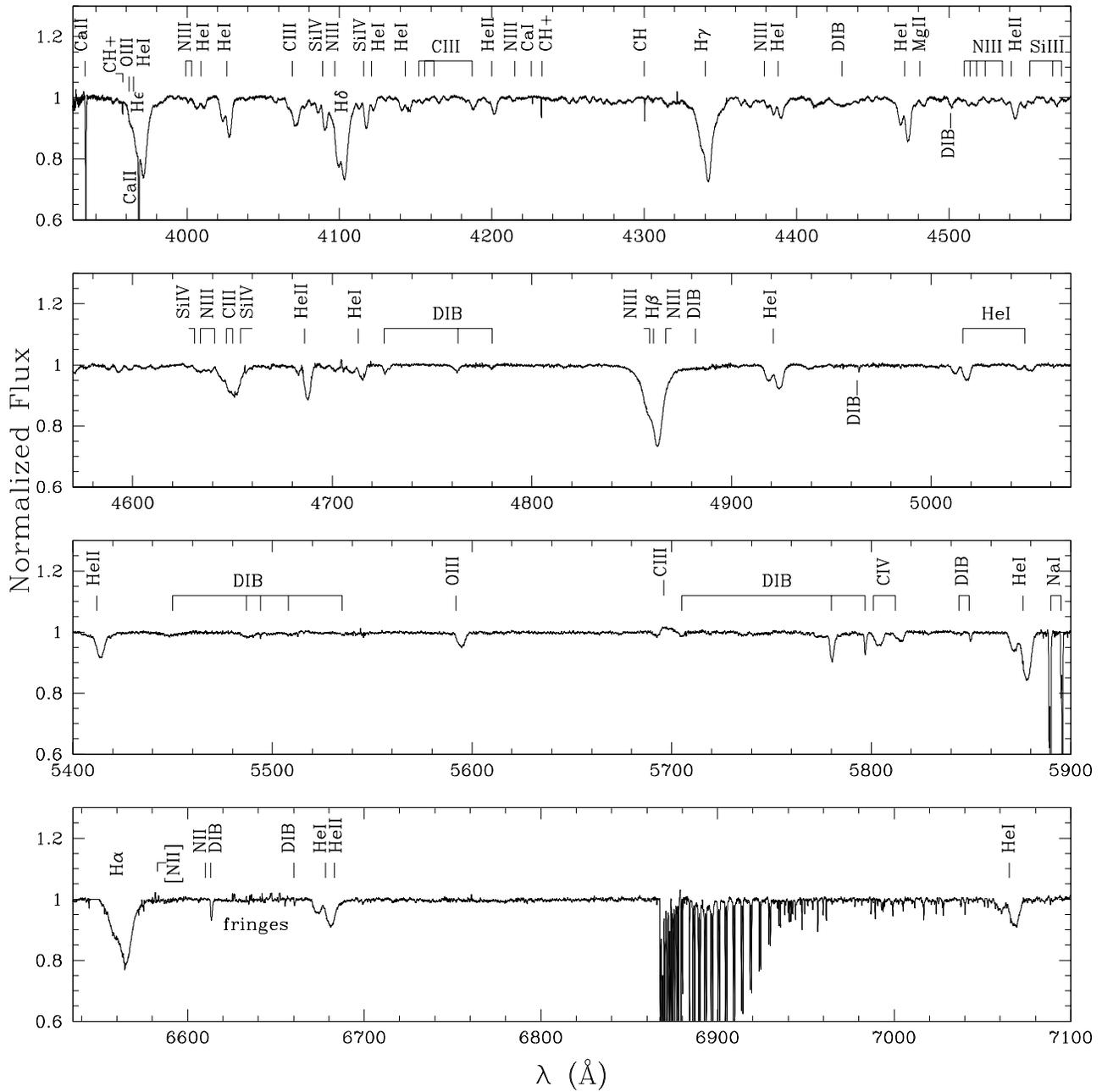}
\caption{Spectrum of \astrobj{HD 152218} as observed with the FEROS spectrograph on HJD\,2\,453\,134.723 ($\phi \sim 0.85$). At this orbital phase, the spectrum of the most massive star is shifted towards the red. The most prominent lines are indicated at their rest wavelengths. }
\label{fig: spec} 
\end{center}
\end{figure*}

\begin{figure}
   \centering
   \includegraphics[width=.5\columnwidth]{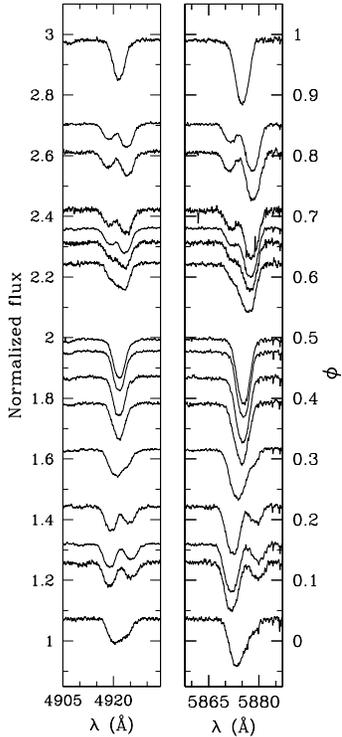}
   \caption{\hea\,\l4922 (left) and \hea\,\l5875 (right) lines at selected phases. The spectra were shifted along the vertical axis according to the value of their phase (right-hand scale). The secondary spectral signature is clearly identified on most spectra but the blended ones.} 
   \label{fig: doppler}  
\end{figure}

The spectrum of \astrobj{HD 152218} is clearly dominated by the usual Balmer, \hea\ and \heb\ lines seen in absorption (Fig. \ref{fig: spec}). Beside these, the spectrum reveals metallic lines of \nc, \sic, \sid, \cd\ and \oc\ that are typical of O-type stars. Only the \cc\,\l5696 line is seen in emission.
 Most of the observed stellar lines are double (see e.g. Fig. \ref{fig: doppler}) and display clear Doppler shifts. The \astrobj{HD 152218} spectrum  also reveals a number of diffuse interstellar bands (DIBs), as well as interstellar absorptions due to \naa, \caa, \cab, CH and CH$^+$. 

We measured the Doppler shifts  by fitting  a sum of Gaussians to the different line profiles. Whenever the components get closer to conjunction, the disentangling of their spectral signature was more difficult. To get reliable results we had to constrain the full width at half maximum (FWHM) of the fitted Gaussian profiles. We thus computed the mean FWHMs of the primary and secondary lines according to the measurements performed near quadrature. Considering the properties of the studied lines, we then decided either to fix the line FWHMs to their mean value or to request  the primary and secondary lines to have equal FWHMs. The resulting radial velocities (RVs) were assigned a lower weight in the determination of the orbital solution (see Sect. \ref{ssect: orb}).
In order to get a reliable solution, we  considered 11 different lines: \hea\,\l\l 4026, 4388, 4471, 4922, 5016, 5876, 7065, \heb\,\l\l\,4542, 4686, 5412 and \oc\,\l5592. Note that the \heb\,\ll4542 and 5412 lines were  rarely deblended and show a faint secondary component only. For these lines, we thus focused on  the primary orbital motion. We also note that the \oc\,\l5592 line is only seen in the primary spectrum. Finally, to compute the RVs associated with the measured Doppler shifts, we adopted the effective wavelengths for O-stars from \citet{CLL77} below 4800 \AA\ and from \citet{Und94} above. % For the metallic lines that are not listed in these latter works, we used the rest wavelengths from \citet{Moore59}. 

% *************************************************************************
% *************************************************************************

% *************************************************************************
% *************************************************************************

% *************************************************************************
\begin{figure}
\centering
\includegraphics[width=0.49\columnwidth]{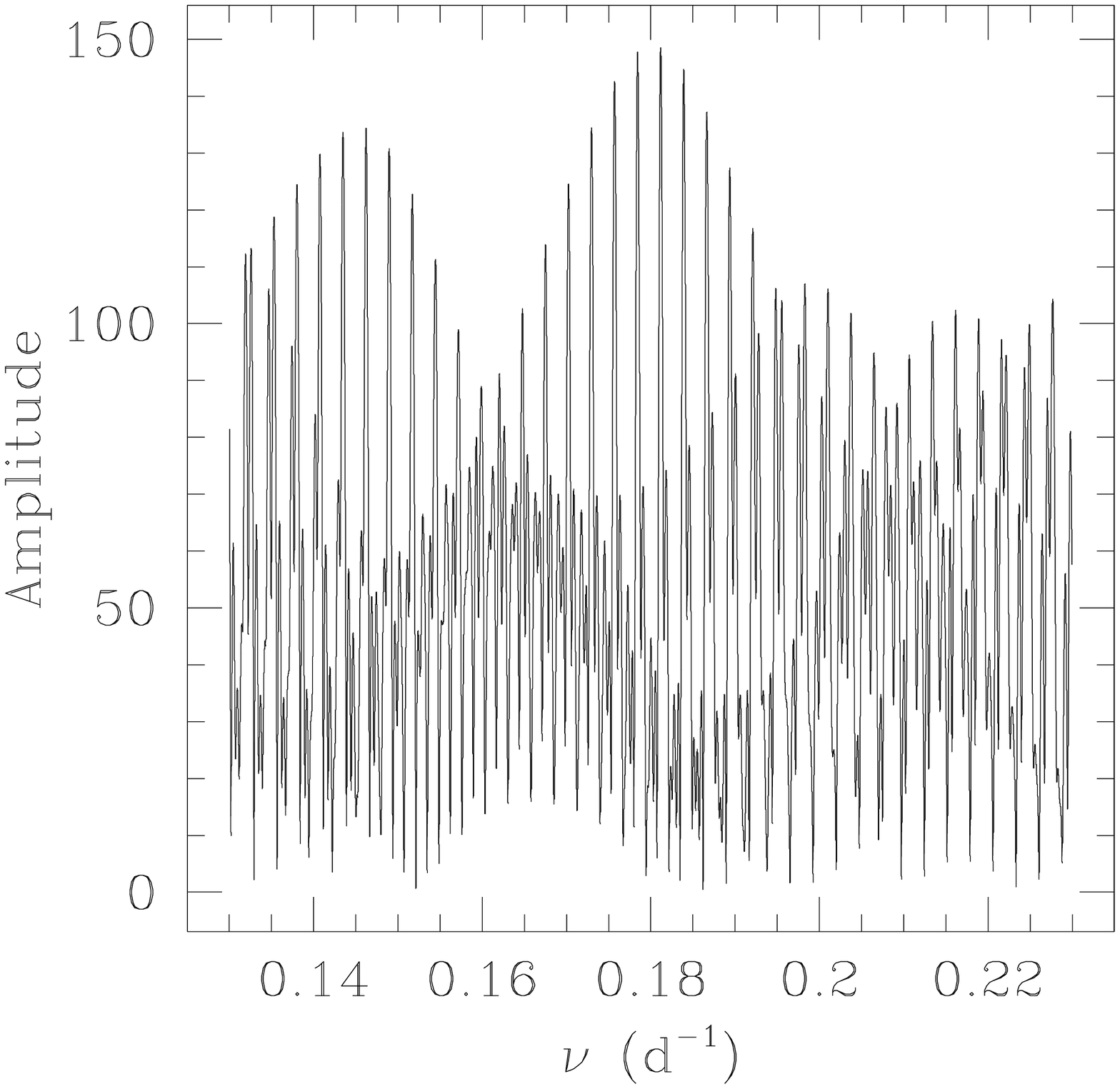}
\includegraphics[width=0.49\columnwidth]{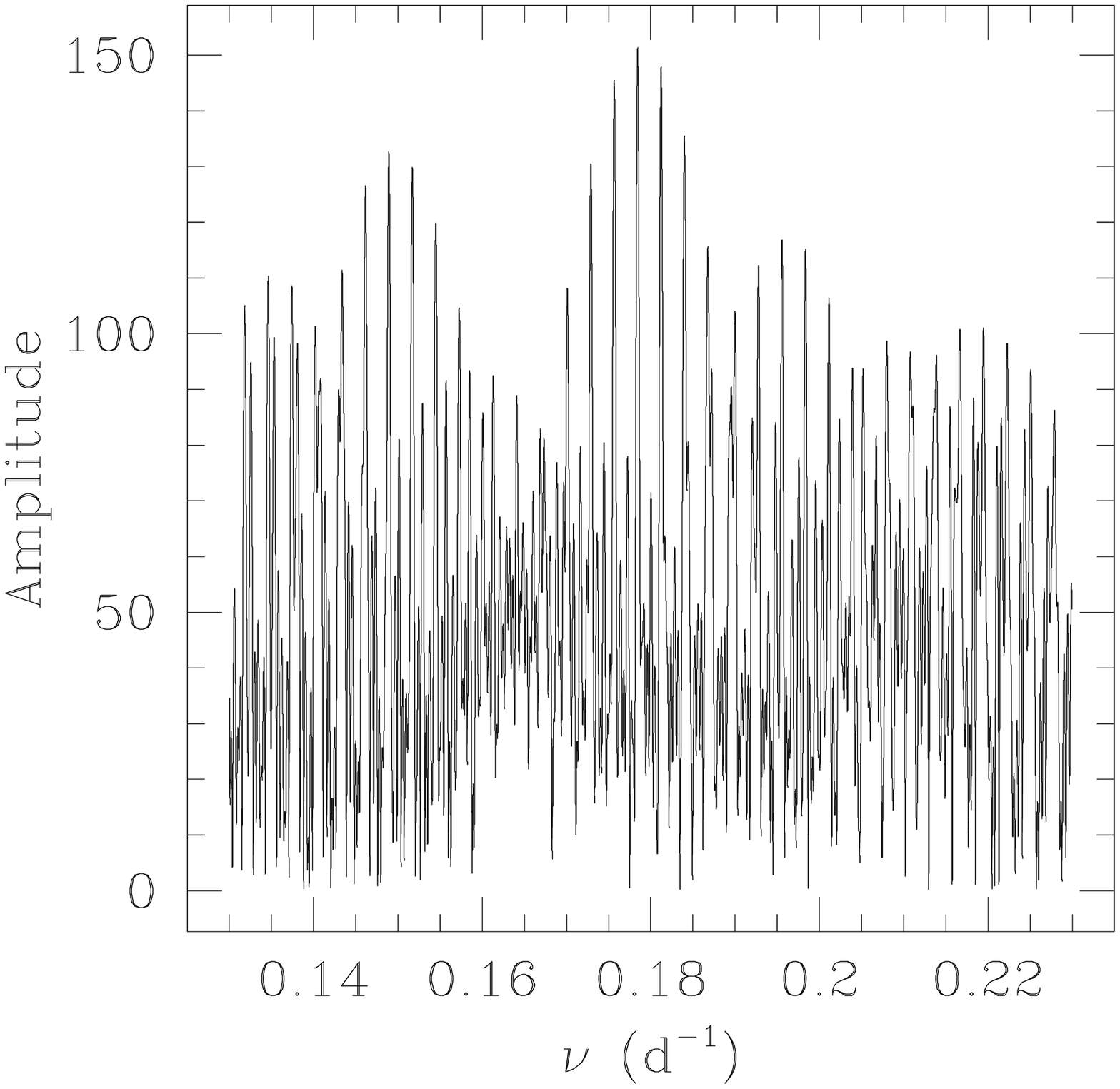}
\caption{{\bf Left panel}: \citetalias{HMM85}  periodogram (i.e. amplitude spectrum) associated with the \fer\ \hea\,\l4471 primary data set. Periodograms associated with other lines are very similar. {\bf Right panel}: idem, using the \ces\ and \bme\ measurements in addition to the \fer\ data. This latter data set favors a period of 5.604~d, corresponding to $\nu \approx 0.1784$~d$^{-1}$. }
\label{fig: alias}
\end{figure}
% *************************************************************************

%_________PERIOD DETERMINATION ____________________________

\subsection{Period determination} \label{ssect: per}

Our period search relies on the method of \citet[ L\&K]{LK65} and on the Fourier-analysis technique of  \citet[\citetalias{HMM85}, see also  \citealt{GRR01} for comments]{HMM85} applied on the different data sets of Table \ref{tab: orbit_indiv}. Using the L\&K algorithm, the obtained periods are in excellent agreement, yielding a mean value of $5.60380\pm 0.00005$~d. The results from the Fourier analysis are more conflicting. The \hea\,\l\,4471 and \heb\,\l\,4686 data sets yield a period of 5.604~days while the other lines indicate 5.519~days. An inspection of the corresponding periodograms revealed that the period determination is strongly affected by a 1-year aliasing, leading to a possible confusion between orbital frequencies of 0.1784 and 0.1812~d$^{-1}$. These latter values correspond respectively to the two period values quoted here above. From Fig.~\ref{fig: alias} (left), it is clear that the period can not be reliably constrained  from the sole \fer\ data. The inclusion of the \ces\ and, to a lesser extent, of the \bme\ data clearly improves the situation (Fig. \ref{fig: alias}, right) but still does  not allow to draw a firm conclusion.

Our data sets consist of spectra spread over 2500~days for the \hea\,\l4471 and \heb\,\l4686 lines  or over 1800~days for the other lines. These observational time bases correspond respectively to natural peak widths $\Delta\nu$ of 0.0004~d$^{-1}$ and 0.0006~d$^{-1}$. The full width at half maximum of the actual peaks in the periodogram is in agreement with these theoretical values. Adopting an uncertainty of one tenth of the peak width, we respectively obtained $\sigma_P= 0.0013$ or 0.0017~d. These values however only account for the uncertainty due to the finite width of the peak, but do not consider the possible confusion between different aliases. \\ 

\begin{figure}
\centering
\includegraphics[width=0.49\columnwidth]{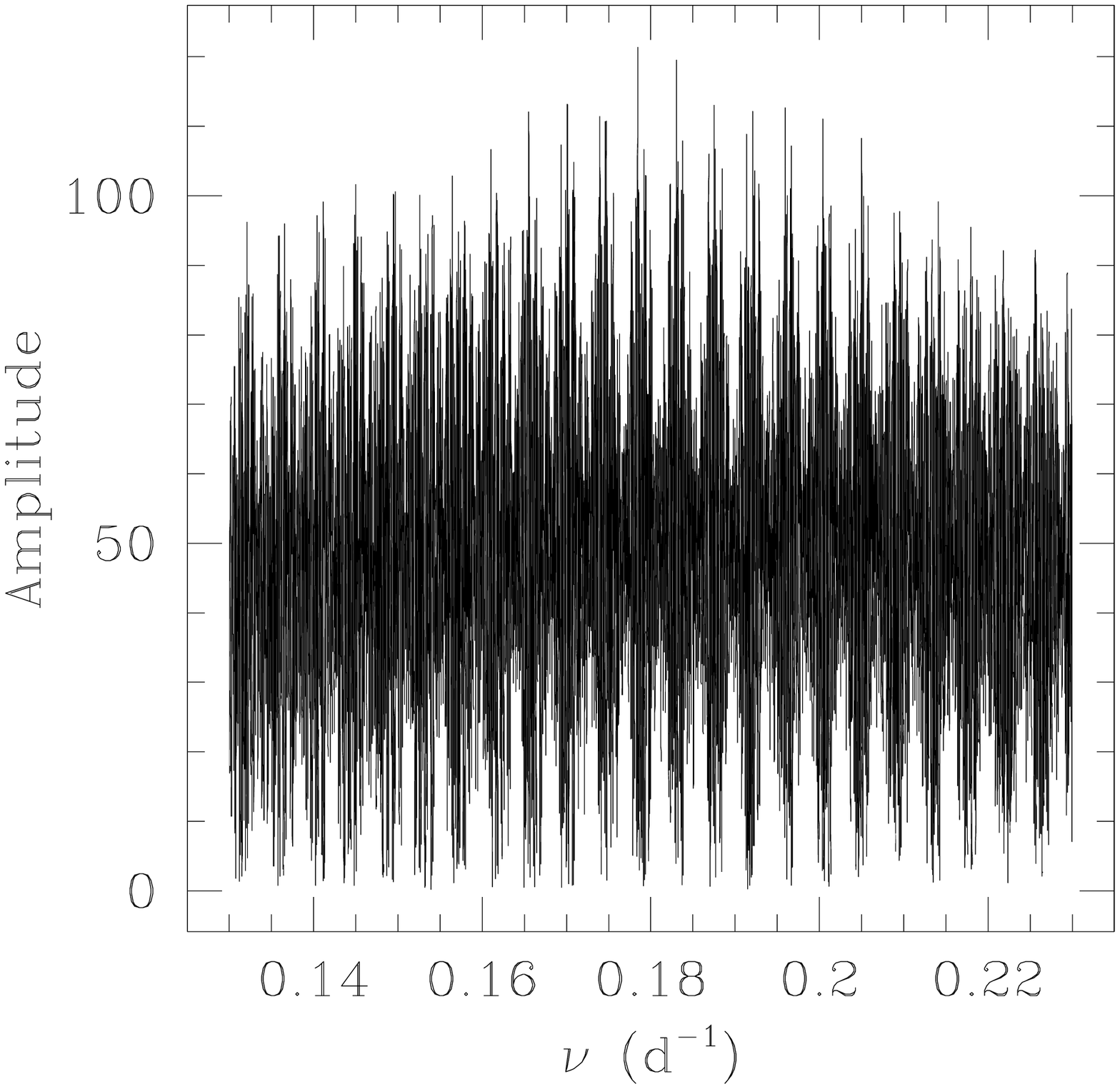}
\includegraphics[width=0.49\columnwidth]{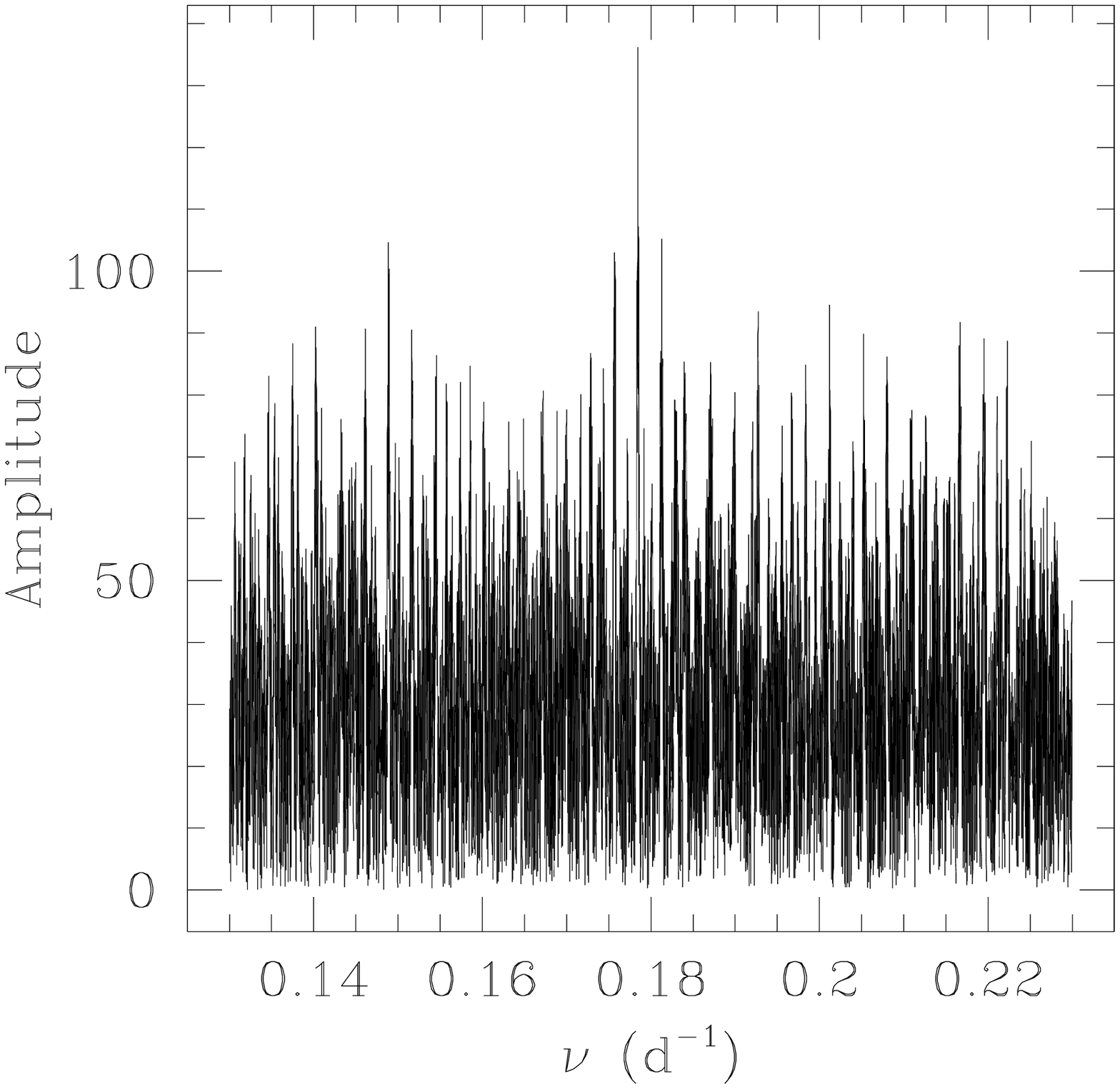}
\caption{{\bf Left panel}: \citetalias{HMM85} periodogram corresponding to the IUE+optical data sets used by \citetalias{SLP97}. {\bf Right panels}: idem, combining our data with published data sets.}
\label{fig: alias2}
\end{figure}

To solve the aliasing problem, we turned to the literature. As \citet[ \citetalias{HCB74}]{HCB74} and \citet[ \citetalias{LM83}]{LM83} derived quite different period values compared to the ones quoted above, we first applied our period search methods using their data sets and found a very large number of aliases covering a frequency range corresponding to period values between about 5 and 6.7~days. Their proposed period values were thus much more poorly constrained than what their error-bars suggested. On the other hand, \citet[\citetalias{SLP97}]{SLP97} obtained a period $P\approx 5.604$~d, very similar to one of our aliases. We used both the L\&K and \citetalias{HMM85} algorithms on the different data sets considered by \citetalias{SLP97} and we re-derived very similar values, except for the related uncertainties. Adopting a typical error corresponding to one tenth of the periodogram peak width, we obtained values about 6 times larger. The periodogram (Fig. \ref{fig: alias2}, left panel) also indicates a  possible confusion with a neighbouring alias at $\nu\approx0.183$~d$^{-1}$. The period value is again not so clearly established. We however note that this latter alias is not present in our data set. By combining the different data sets, we  could thus hope to definitely constrain the period. 

% *************************************************************************
% *************************************************************************

% *************************************************************************
\begin{sidewaystable*} 
\begin{center}
\caption{Orbital solutions for \astrobj{HD 152218} as derived from different lines. The usual notations have been adopted. $s_y/s_x$ gives the adopted secondary to primary ratio of the uncertainties on the measured RVs (see Sect. \ref{ssect: orb}). The quoted error-bars are the 1-$\sigma$ uncertainties. }
\label{tab: orbit_indiv}
\begin{tabular}{l c c c c c c c c c}
\hline
Lines     &  $P$  & $s_y/s_x$ & $e$            & $\omega$        &   $K_1$         & $K_2$           & $\gamma_1$      &  $\gamma_2$     &  r.m.s.  \\
          &  (days)&          &                & (\degr)         &  (\kms)         & (\kms)          & (\kms)          &  (\kms)         & (\kms)\\
\hline
\hea\,\l4026 & 5.60400 & 1.4 & $0.268 \pm 0.007$ & $104.4 \pm 1.7$ & $164.4 \pm 1.3$ & $209.2 \pm 1.6$ & $-16.7 \pm 1.4$ & $-17.3 \pm 1.6$ &  5.6 \\ 
\hea\,\l4388 & 5.60395 & 1.2 & $0.253 \pm 0.011$ & $ \hspace*{1.7mm}99.4 \pm 3.3$ & $159.2 \pm 2.2$ & $208.1 \pm 2.8$ & $-11.4 \pm 2.3$ & $-15.5 \pm 2.7$ &  9.5 \\ 
\hea\,\l4471 & 5.60386 & 1.7 & $0.242 \pm 0.008$ & $101.7 \pm 2.6$ & $160.2 \pm 1.6$ & $215.0 \pm 2.2$ & $-26.4 \pm 1.7$ & $-30.6 \pm 2.1$ &  7.7 \\ 
\heb\,\l4686 & 5.60391 & 1.7 & $0.275 \pm 0.008$ & $108.4 \pm 1.8$ & $165.8 \pm 1.7$ & $207.9 \pm 2.2$ & $-15.4 \pm 1.7$ & $-16.8 \pm 1.9$ &  7.1 \\ 
\hea\,\l4922 & 5.60393 & 1.4 & $0.261 \pm 0.008$ & $103.0 \pm 2.1$ & $159.8 \pm 1.4$ & $214.4 \pm 1.8$ & $-18.0 \pm 1.6$ & $-21.8 \pm 1.9$ &  6.3 \\ 
\hea\,\l5016 & 5.60405 & 1.6 & $0.269 \pm 0.010$ & $102.6 \pm 2.4$ & $167.4 \pm 2.0$ & $215.2 \pm 2.6$ & $-16.3 \pm 2.0$ & $-17.2 \pm 2.4$ &  8.3 \\ 
\hea\,\l5876 & 5.60410 & 1.6 & $0.271 \pm 0.006$ & $104.5 \pm 1.5$ & $162.4 \pm 1.1$ & $221.4 \pm 1.5$ & $-21.9 \pm 1.2$ & $-21.1 \pm 1.5$ &  4.8 \\ 
\hea\,\l7065 & 5.60410 & 1.7 & $0.255 \pm 0.008$ & $101.0 \pm 2.2$ & $165.1 \pm 1.6$ & $214.0 \pm 2.1$ & $-17.1 \pm 1.7$ & $-11.7 \pm 2.0$ &  6.4 \\ 
\hline
\heb \l4542 & 5.60400 &  -- & $0.288 \pm 0.014$ & $101.3 \pm 2.2$ & $160.9 \pm 2.1$ &   --            & $-21.1 \pm 1.2$ &   --            &  6.5   \\
\heb\,\l5412 & 5.60419 &  -- & $0.302 \pm 0.011$ & $104.0 \pm 1.6$ & $161.4 \pm 1.7$ &   --            & $-16.5 \pm 1.0$ &   --            &  5.2   \\
 \oc\,\l5592 & 5.60409 &  -- & $0.274 \pm 0.012$ & $106.1 \pm 2.0$ & $162.7 \pm 1.9$ &   --            & $-17.1 \pm 1.7$ &   --            &  6.0   \\
\hline
\end{tabular}
\end{center}
\end{sidewaystable*}
% *************************************************************************

To combine our data with other observational data sets, we first averaged the RV measurements obtained from the different lines quoted in Table \ref{tab: orbit_indiv}\footnote{ In this first approach, we did not correct the individual line RVs for their systemic velocities but we rather performed a direct average. Indeed the systemic velocities should be obtained from RV curve fitting. They thus require an first accurate estimate of the period}. Then, applying the L\&K and \citetalias{HMM85} period searches to all the available RV measurements, we derived values of $P= 5.6039$ and 5.6040~d respectively. The corresponding Fourier periodogram is displayed in Fig. \ref{fig: alias2} (right) and clearly shows that the orbital peak is now well isolated. The time span of the whole observational set is $\sim$21~915~d, yielding a theoretical periodogram peak width of $\Delta \nu=4.6\times 10^{-5}$~d$^{-1}$. The observed width of the peak agrees with this value and yields $\Delta P = 1.4\times 10^{-3}$~d. We adopted a final uncertainty corresponding to one tenth of the peak width, i.e.\ $\sigma_P=1.4\times 10^{-4}$~d.

%___________________ ORBITAL ELEMENTS _______________________________

\subsection{Orbital elements} \label{ssect: orb}

Now that the period is reliably constrained, we can derive the orbital parameters of the system by fitting the observed RV-curves assuming a Keplerian motion for the two components. We first computed a series of orbital solutions using the RV sets associated with the different absorption lines listed in Table \ref{tab: orbit_indiv}. For SB1 lines, we used the algorithm of \citet{WHS67} in its original form and assigned the same weight to each measurement. For the data sets associated with SB2 lines, we used a modified version of the algorithm, adapted to SB2 systems as explained in \citet{SGR06_219}, and we only considered those data points for which the two components could be measured with a reasonable accuracy. We assigned a half weight to those RV points for which the FWHMs had to be constrained (see Sect. \ref{ssect: RVs}). Because of the lower SNR of the \bme\ data, we gave them a  weight of one fourth while computing the orbital solution. In the observed wavelength domain, the \bme\ measurements were further not included for the fainter lines nor for the lines more difficult to disentangle (\hea\,\ll4388, 5016, \heb\,\l4542, \oc\,\l5592).
In our derived solutions,  we adopted a period value and a relative primary to secondary weight ratio ($s_y/s_x)$ that yield  the lowest $\chi^2$.  These best solutions, including a reappraisal of the period\footnote{The period has been adjusted simultaneously together with the other orbital parameters, adopting the values derived in Sect.~\ref{ssect: per} as initial guess.},  are presented in Table \ref{tab: orbit_indiv} together with the corresponding root-mean-square (r.m.s) residuals of the fit. 

\begin{figure}
   \centering
   \includegraphics[width=.9\columnwidth]{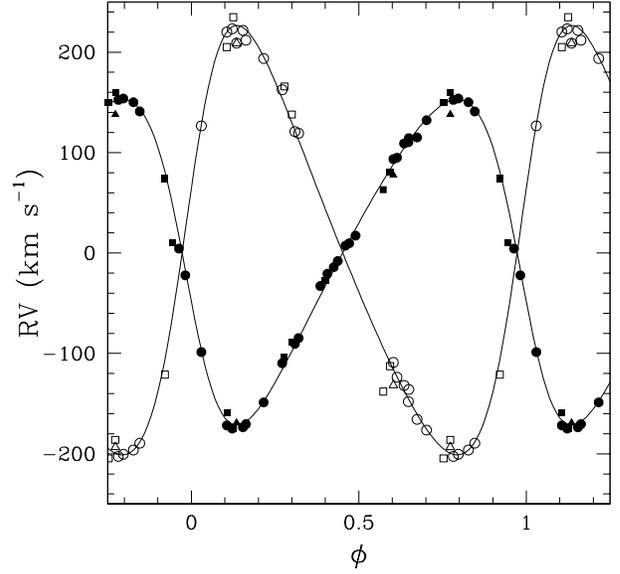}
      \caption{\astrobj{HD 152218} RV-curves  corresponding to the \hea\ solution of Table \ref{tab: orbit}. Note that the respective systemic velocities have been subtracted. RV measurements listed in Table \ref{tab: opt_diary} have been overplotted. Filled and open symbols are for the primary and secondary components respectively. Different symbols indicate different instruments used: \ces, squares; \bme, triangles; \fer, circles.}
         \label{fig: sb2}
   \end{figure}

In a next step, we computed the mean RVs of the SB2 \hea-lines quoted in Table \ref{tab: orbit_indiv} as well as of the sole primary lines. For this purpose, we first shifted the individual RVs to a common frame, by subtracting the individual systemic velocities. The resulting mean RVs are listed in Table \ref{tab: opt_diary}. The orbital solutions computed using the averaged primary (SB1) and the averaged \hea-line (SB2) RV measurements are given in Table \ref{tab: orbit} together with the physical parameters of the system. Fig. \ref{fig: sb2} displays the RV curves corresponding to the averaged SB2 solution. The latter solution  will be adopted throughout the rest of the paper to constrain the physical properties of the \astrobj{HD 152218} components (see Sect. \ref{sect: physic}). Corresponding primary and secondary apparent systemic velocities, as obtained from a weighted mean of the values quoted in Table \ref{tab: orbit_indiv}, are $\overline{\gamma_1}=-18.7\pm0.6$~\kms\ and $\overline{\gamma_2}=-19.2\pm0.7$~\kms.

Our final SB2 solution is significantly different from the most recent determination by \citet[ \citetalias{GM01}]{GM01} who obtained $P\sim4.9$~d and $e\sim0.4$. We however mostly recover, within the error-bars,  the results of an older study  by \citetalias{SLP97} Indeed, our new values fall within 3-$\sigma$ from the results of \citetalias{SLP97}, though with a slightly smaller eccentricity and larger RV-curve semi-amplitudes. Only the longitude of the periastron seems to differ significantly: $80\fdg6\pm2\fdg1$ for \citetalias{SLP97} versus $104\fdg2\pm1\fdg6$ for our solution. This suggests the presence of an apsidal motion of the order of 20\degr\ over about 7~yr. This question is addressed in the next section.

\begin{table}
\centering
\caption{Orbital and physical parameters of \astrobj{HD 152218} as deduced from two different data sets: the averaged RVs computed over selected primary SB1 lines (Col. 2) and over \hea\ SB2 lines (Col. 3). The usual notations have been adopted. $T_0$ is the time of periastron passage and is adopted as phase zero (i.e.\ $\phi=0.0$). Note that the primary and the \hea\ lines solution were computed in the systemic velocity frame.}
\label{tab: orbit}
\tiny
\begin{tabular}{c c c}
\hline
                       & Prim.            & \hea\ lines          \\
\hline
$P$ (d)                & 5.60404          &   5.60391           \\     
$s_y/s_x$              &  n.              &   1.5               \\
$m_1/m_2$              &  n.              &   1.319 $\pm$ 0.014 \\
$e      $              & 0.275 $\pm$ 0.008&   0.259 $\pm$ 0.006 \\
$\omega $ (\degr)      & 104.2 $\pm$ 1.6  &   104.0 $\pm$ 1.6   \\
$T_0$ (HJD             & 3196.493         &    3197.229         \\
$-$2\,450\,000)        &\hspace*{3mm}$\pm$0.024 & \hspace*{3mm} $\pm$ 0.025 \\
$K_1$\ (\kms)          & 162.4 $\pm$ 1.1  &   162.2 $\pm$ 1.2    \\
$K_2$\ (\kms)          &  n.              &   213.9 $\pm$ 1.5    \\
$\gamma_1$\ (\kms)     &    0.7 $\pm$ 0.9 &     0.9 $\pm$ 1.2    \\
$\gamma_2$\ (\kms)     & n.               &  $-$0.6 $\pm$ 1.4    \\
$a_1 \sin i$ (\rsol)   & 17.28 $\pm$ 0.13 &   17.34 $\pm$ 0.13   \\
$a_2 \sin i$ (\rsol)   &  n.              &   22.87 $\pm$ 0.17   \\
$m_1 \sin^3 i$ (\msol) &  n.              &   15.82 $\pm$ 0.26   \\
$m_2 \sin^3 i$ (\msol) &  n.              &   12.00 $\pm$ 0.19   \\
%$R_{RL,1}/a$          &                  & 0.4031  $\pm$ 0.0009 \\
%$R_{RL,2}/a$          &                  & 0.3552  $\pm$ 0.0009 \\
r.m.s. (\kms)          & 4.1              & 6.1                  \\
\hline
\end{tabular}
\end{table}

%________________________ APSIDAL MOTION _________________________________

\subsection{Apsidal motion} \label{ssect: orb_lit}

As already mentioned, several authors have collected a number of spectroscopic observations of \astrobj{HD 152218} and measured the RVs of its two components. The largest sets are due to \citet{Str44}, \citetalias{HCB74}, \citetalias{SLP97} and \citetalias{GM01} who respectively obtained 5, 15, 22 and 7 spectra. \citeauthor{Str44} published the observing dates using universal times with day and month. \citetalias{HCB74} later presented corresponding Julian dates (JD). We however note that, though quoted otherwise, these dates are not heliocentric Julian dates (HJD) but simply JD. We thus recomputed the appropriate HJD before using the \citeauthor{Str44} data set. A word of caution is also required for the \citetalias{GM01} observations. As already mentioned above, \citetalias{GM01} quoted erroneous HJDs for part of their observations\footnote{In \citetalias{GM01}'s Table 2, the HJDs around 2~449~910 are probably correct, those around 2~450~590 are overestimated by exactly one day for observations obtained after midnight UT. We could not check the more recent observations obtained by \citetalias{GM01} at HJDs$\sim$2~451~360.}. We thus corrected the corresponding values prior to the inclusion of  the \citetalias{GM01} observations. We  completed these four sets by an additional measurement from \citet{CLL77} and another one by \citetalias{LM83}. We finally added the RVs from the present work (obtained as the average of the $\gamma$-corrected individual lines measurements) and we adopted the systemic velocities derived in Sect. \ref{ssect: orb}.

\begin{figure}
\centering
\includegraphics[width=.9\columnwidth]{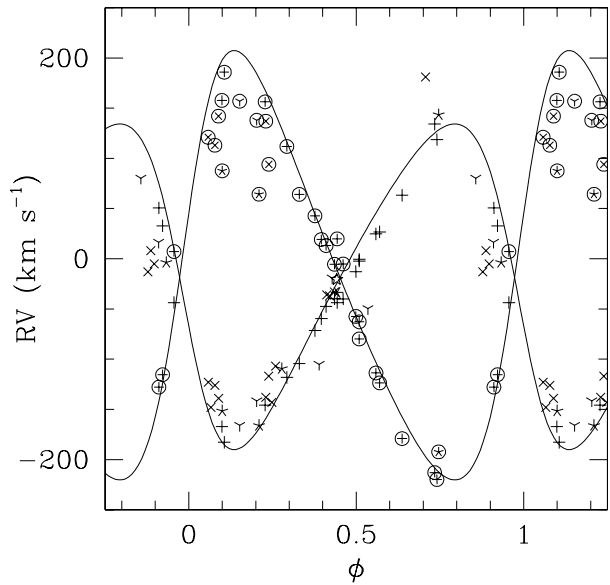}
\caption{Literature RV measurements plotted over the \hea\ SB2 RV-curves of Fig. \ref{fig: sb2}, shifted according  to their respective apparent systemic velocities. The phases were computed using the \hea\ ephemeris of Table \ref{tab: orbit}. Different symbols indicate different works: (five branch star) \citet{Str44}; ($\times$) \citetalias{HCB74}; (asterisk) \citetalias{LM83}; (open five-branch star) \citet{CLL77}; ($+$) \citetalias{SLP97}; (Y) corrected \citetalias{GM01}. These symbols inscribed in open circles indicate secondary RV measurements.}
\label{fig: litover}
\end{figure}

\begin{figure}
\centering
\includegraphics[width=.9\columnwidth]{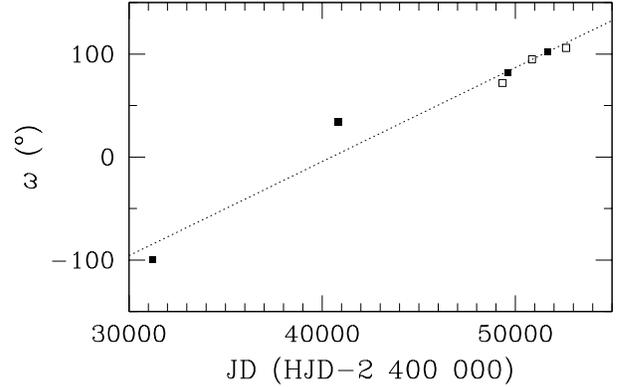}
\caption{Values of the argument of periastron $\omega$ computed for different epochs. The dotted line gives the best-fit relation, corresponding to an apsidal motion of $9\fdg1\pm0\fdg8 \times10^{-3}$~d$^{-1}$, or about $3\fdg3\pm0\fdg3$~yr$^{-1}$.}
\label{fig: aps}
\end{figure}

In a first approach, Fig. \ref{fig: litover} shows the literature measurements plotted over the \hea\ orbital solution of Table \ref{tab: orbit}. The secondary measurements present a larger scatter and we will thus focus on the primary data for the rest of this section. In Fig. \ref{fig: litover}, we also observe systematic shifts between the literature measurements and our newly derived orbital solution. As suggested in the previous section, this most probably comes from the presence of an apsidal motion. Fixing the period and the eccentricity at the values quoted in Table \ref{tab: orbit} for the \hea-solution, we adjusted the different data sets of Table  \ref{tab: aps} and derived, for each epoch, the best-fit longitude of periastron. These values are plotted in Fig. \ref{fig: aps}.  We note that, using the  RV sets of \citetalias{HCB74} and \citet{SLP97}, we re-derived the longitude values respectively quoted in their proposed orbital solutions. 
We also built three subsets of data using the data of \citetalias{GM01}, \citetalias{SLP97} and ours. These sets, labelled A, B and C in Table \ref{tab: aps}, correspond respectively to data obtained before HJD~=~2~450~000, in-between HJD~=~2~450~000 and HJD~=~2~451~800 and after HJD~=~2~452~000.  These sets are however not independent from the sets labelled `SLP97' and `This work' in Table \ref{tab: aps} as they are formed by the same observations. We consider our approach to provide a kind of moving average over the time. It can also be seen as a way to give more weight to the more recent data, that probably provide a better accuracy than the older sets of \citeauthor{Str44} and \citetalias{HCB74}. Using a linear regression, we then computed the best-fit linear relation that reproduces the observed values, the slope of which provides the rate of the apsidal motion. We obtained: $\dot{\omega} = 9\fdg1 \pm 0\fdg8 \times10^{-3}$~d$^{-1}$, corresponding to about $\dot{\omega} = 3\fdg3 \pm 0\fdg3$~yr$^{-1}$. It is worth to note here, that, because of the similar $K_1$ and $K_2$, the shape of the RV-curve is little altered  by a change of $\omega$ of $\pi$. By chance, thus, neither the \citet{Str44} data nor those of \citetalias{HCB74} could strongly bias the orbital period determination of Sect. \ref{ssect: per}. Indeed similar results are obtained while including or rejecting these two sets from the period search analysis.

A word of caution is however necessary concerning the rate of the apsidal motion. Indeed, we have also performed a similar analysis using our data only. These were grouped into five different subsets according to the different observing epochs. The apsidal motion is again clearly detected, even considering the comparatively smaller time span of our spectroscopic campaign. The obtained rate is however somewhat slower: $\dot{\omega} = 1\fdg4 \pm 0\fdg6$~yr$^{-1}$. Accounting in addition for the \citetalias{SLP97} data, we rather obtain $\dot{\omega} = 3\fdg0 \pm 0\fdg6$~yr$^{-1}$. Although there is little doubt left about the existence of an apsidal motion, the {\it exact} rate at which the latter occurs might still be uncertain. This probably reflect the difficulty to combine inhomogeneous data sets.

\begin{table}
\centering
\caption{Best fit values for the longitude of the periastron $\omega$ at different epochs. The data sets labelled `A' to `C' are described in the text.}
\label{tab: aps}
\begin{tabular}{r r r r }
\hline
\vspace*{-3mm}\\
Data set & $\overline{\mathrm{HJD-2~400~000}}$ & $\omega$ (\degr) & r.m.s. (\kms) \\
\hline
\citet{Str44}                   & 31223.5 & $-100$ &  5.4 \\
\citetalias{HCB74}              & 40818.9 &    34  & 11.7 \\
\citetalias{SLP97}~$^\mathrm{a}$ & 49627.5 &    82  &  7.6 \\
This work                       & 51652.1 &   102  &  6.6 \\
\hline
A & 49333.8 & 72 & 19.9 \\
B & 50861.8 & 95 & 11.8 \\
C & 52621.3 &106 &  4.2 \\
\hline
\end{tabular}\\
a. The first point of the \citetalias{SLP97} set is separated by about 3300~days from the other points of the set and was not included in the fit.\\
\end{table}

%_____________________ PHYSICAL PARAMETERS ____________________________

\section{Physical parameters} \label{sect: physic}

\subsection{Spectral types and luminosity classes } \label{ssect: spt}

The dominant  spectral signature of the primary component in the spectrum of \astrobj{HD 152218} is characteristic of a late O-type star. We adopted the classification criteria from \citet{Con73_teff} as adapted to late O-stars by \citet{Mat88}. They are based on the equivalent width (EW) ratio of the \hea\,\l4471 and \heb\,\l4542 lines. Doing this, we only considered the EWs measured on the spectra where the two components are disentangled.  For the primary, we obtained a mean $ \log W'_1 = \log  W_{\lambda4471}- \log W_{\lambda4542}=0.33\pm0.06$ which corresponds to a spectral type O9, with spectral type O8.5 within $1\,\sigma$. For the secondary, we obtained $ \log W'_2=0.67\pm0.08$. This indicates an O9.7 star, with the O9.5 type at 1-$\sigma$.

To determine the luminosity classes, we adopted the criterion from \citet{CA71} based on the EW ratio of the \sid\,\l4089 and \hea\,\l4144 lines. We respectively obtained $\log W''_1 = \log W_{\lambda4089} - \log W_{\lambda4144}=0.25\pm0.05$ and $\log W''_2=-0.12\pm0.13$. These values  lead to giant and main sequence luminosity classes for the primary and secondary components respectively. 
 According to \citet{Mat88}, we also measured $\log W'''_1= \log W_{\lambda4388} + \log W_{\lambda4686}=5.03\pm0.04$. This corresponds to a  supergiant if the primary brightness ratio $l_1=\frac{L_1}{L_\mathrm{tot}}>0.76$, to a giant if $0.76>l_1>0.43$ and to a main sequence star otherwise.
For the secondary, we get $\log W'''_2=4.22\pm0.15$. The respective critical values for $l_2=\frac{L_2}{L_\mathrm{tot}}$ are this time 0.12 and 0.07. Such a low contribution of the secondary to the total flux is improbable. \citeauthor{Mat88} criterion thus indicates a supergiant secondary, which is clearly at odds with the result of the \citet{CA71} criterion. However, the \citeauthor{Mat88} criterion can be biased if the \heb\l4686 line is partly filled by emission produced e.g.\ in a wind interaction region, which is possibly the case in HD~152218 (see Sect.~\ref{sect: xmm})
The comparison of our spectra with the atlas of standards \citep{WF90} seems to favor a giant class for the primary star, but it is very difficult to reject the main sequence option. It also indicates that the secondary is most probably a main sequence star. \\

%_______________________ BRIGHNTESS RATIO & EVOLUTIONARY STATUS ______________

\subsection{Optical brightness ratio and evolutionary status} \label{sect: bright}

We roughly estimated the optical brightness ratio based on the dilution of the primary and secondary lines in the spectrum of \astrobj{HD 152218}. For this purpose, we compared the mean EWs of the  primary and secondary lines with typical (averaged) EWs of respectively O9 \citep{CA71, Con73_nlte, Mat88, Mat89} and O9.7 stars \citep{Mat88, Mat89}. We first compared the primary line strength with typical O9III EWs. Based on the \hea\,\l\l4026, 4144, 4388, 4471, \heb\,\l\l4542, 4686 and \sid\,\l4088 lines, we obtained  an averaged brightness ratio of $l_1=0.76\pm0.05$. Using typical O9V EWs yields a lower value $l_1=0.68\pm0.11$. We however note that the \sid\,\l4088 line is particularly strong in the primary spectrum and is the main cause for the large scatter obtained. Rejecting the \sid\,\l4088 line from the comparison with O9V typical EWs, we obtained $l_1=0.63\pm0.04$. 

\begin{figure}
\centering
\includegraphics[width=.9\columnwidth]{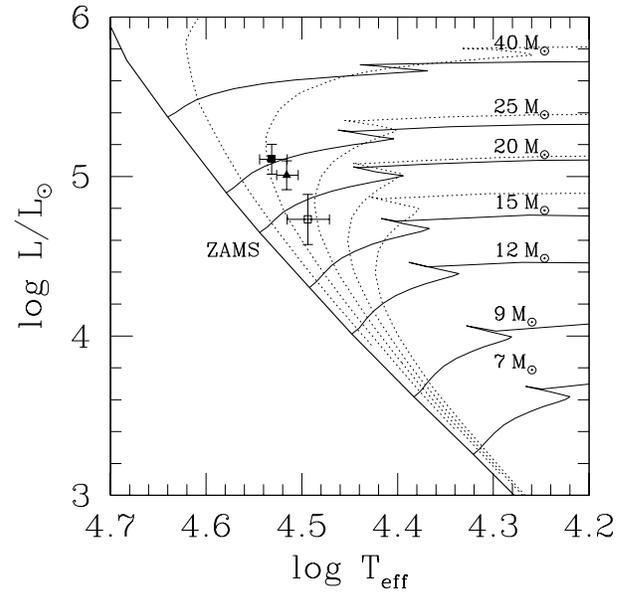}
\caption{\astrobj{HD 152218} primary (filled symbols) and secondary (open symbol) location in the H-R diagram. Triangle (resp. square) symbols indicate an adopted giant (resp. main sequence) luminosity class. The evolutionary tracks from \citet{SSM92} have been overplotted (plain lines) together with isochrones (dotted lines) computed for ages ranging from 2 to 10 Myr with a step of 2 Myr. }
\label{fig: hr}
\end{figure}

Turning to the secondary component, we also compared the secondary line strengths to the typical EWs reported by \citet{Mat88, Mat89}, who only  considered the \hea\,\ll4388, 4471 and \heb\,\ll4542, 4686 lines. We note that the \heb\,\l4686 line provides results that clearly differ from the three other lines, but this is not surprising since the \heb\,\l4686 line is particularly faint in the secondary spectrum. Mean values for the three other lines indicate $l_2=0.34\pm0.02$ and $0.41\pm0.05$ for dwarf and giant star respectively. The obtained values for the primary and secondary contribution to the observed fluxes are thus in acceptable agreement with each other. They also agree with the results of \citet{PBG94} who quoted a primary to secondary UV flux ratio of 2.0. In the following, we will adopt a primary to secondary brightness ratio corresponding to $l_1=0.67 \pm0.10$, which reasonably well reproduces the results of the above discussion. 

\subsection{Effective temperatures and radii}

Sung (2005, private communication) reported $V=7.562$ and $B-V=0.191$ for the system. Using $R=3.3$ \citep{SBL98}, adopting the absolute colors quoted by  \citet{SK82} and a distance modulus $DM=11.07\pm0.04$ \citep[ and references therein]{SGR06}, preliminary computations indicate that the system is too faint to harbour a giant primary, unless its contribution $l_1$ to the observed flux brightness were very close to unity. This is clearly not confirmed by the previous discussion and we suggest that the primary in \astrobj{HD 152218} could be a class IV object. Correcting from the reddening and the distance effects, we obtained $M_\mathrm{V, 1} =  -4.72 \pm 0.22$ and $M_\mathrm{V, 2} = -3.95 \pm 0.37$. Because the intrinsic colour $(B-V)_0$ of an O-type star does almost not depend on the stellar gravity, we note that these values are free from the hypotheses on the luminosity class. According to \citet{HM84}, these values correspond to the primary being a V-IV star and the secondary, a V star. Adopting the \citeauthor{HM84} temperature scale and bolometric corrections, we derived $M_\mathrm{bol, 1}=-8.0\pm0.2$ (resp. $M_\mathrm{bol, 1}=-7.8\pm0.2$) if the primary is a dwarf (resp. giant) star. We obtained $M_\mathrm{bol, 2}=-7.1\pm0.4$ for the main sequence secondary. Finally, using temperatures of $T^\mathrm{O9III}_\mathrm{eff}=32.8^{33.6}_{31.9}$~kK (resp. $T^\mathrm{O9V}_\mathrm{eff}=34.0^{35.0}_{32.8}$~kK), we derived a primary radius of $10.3\pm1.3$~\rsol\ (resp. $9.9\pm1.2$~\rsol). For the secondary we interpolated the \citeauthor{HM84} temperature scale to obtain values of $T^\mathrm{O9.7V}_\mathrm{eff}=31.2^{32.6}_{29.8}$~kK. This corresponds to $R_2=7.9\pm1.7$~\rsol, where the quoted confidence intervals have been computed by adopting the parameters of the neighbouring spectral sub-types.

%\begin{figure}
%\centering
%\includegraphics[bb=18 144 570 560,width=\columnwidth,clip]{hd218hr1}
%%\includegraphics[bb=18 144 570 560,width=\columnwidth,clip]{hd218hr2}
%\caption{X-ray hardness ratio of \astrobj{HD 152218} plotted versus the phase.  The adopted definition is given on top of the panel. }
%\label{fig: xhr}
%\end{figure}

\begin{figure*}
\centering
\includegraphics[width=.4\textwidth]{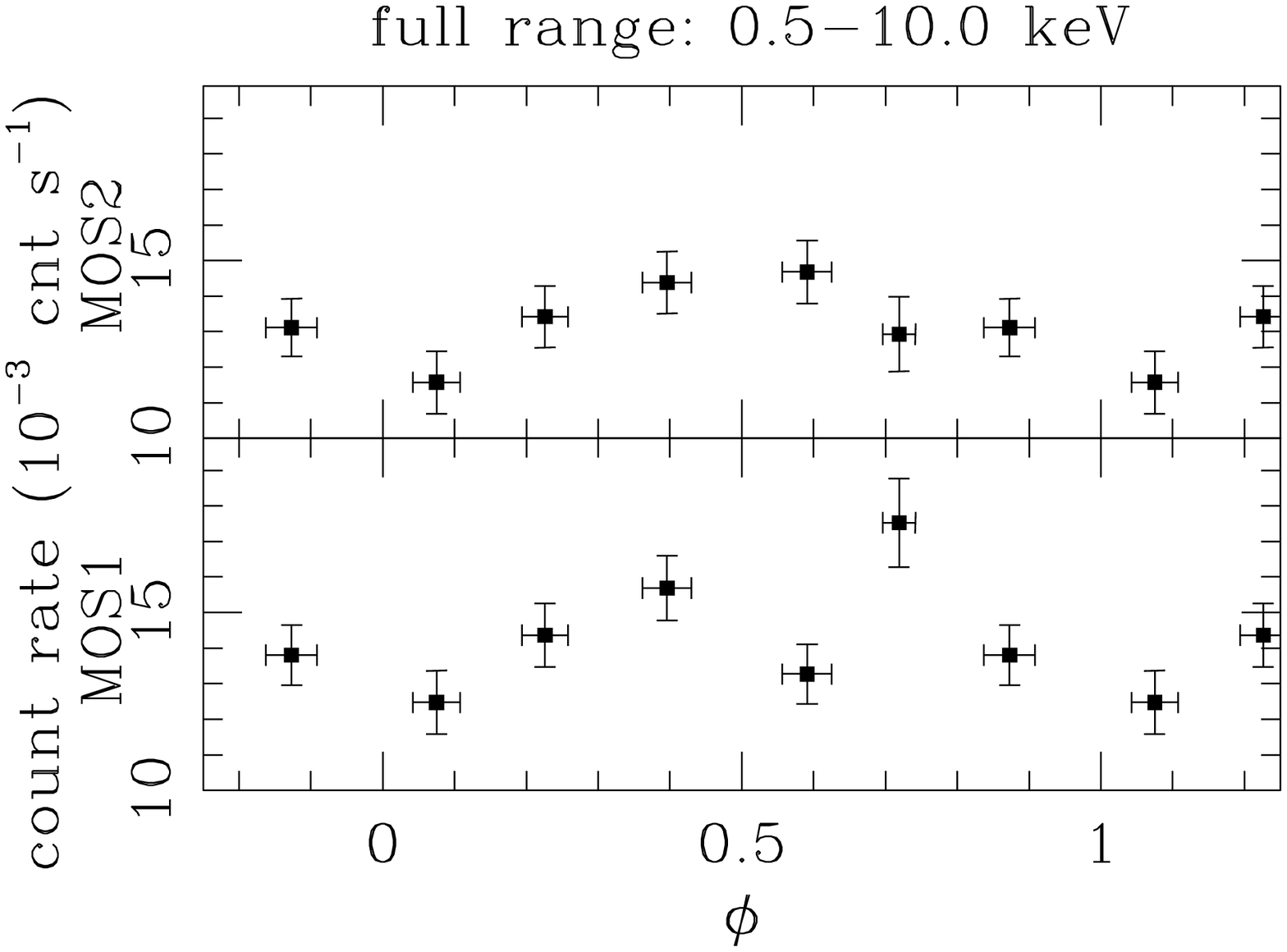}\hspace*{5mm}
\includegraphics[width=.4\textwidth]{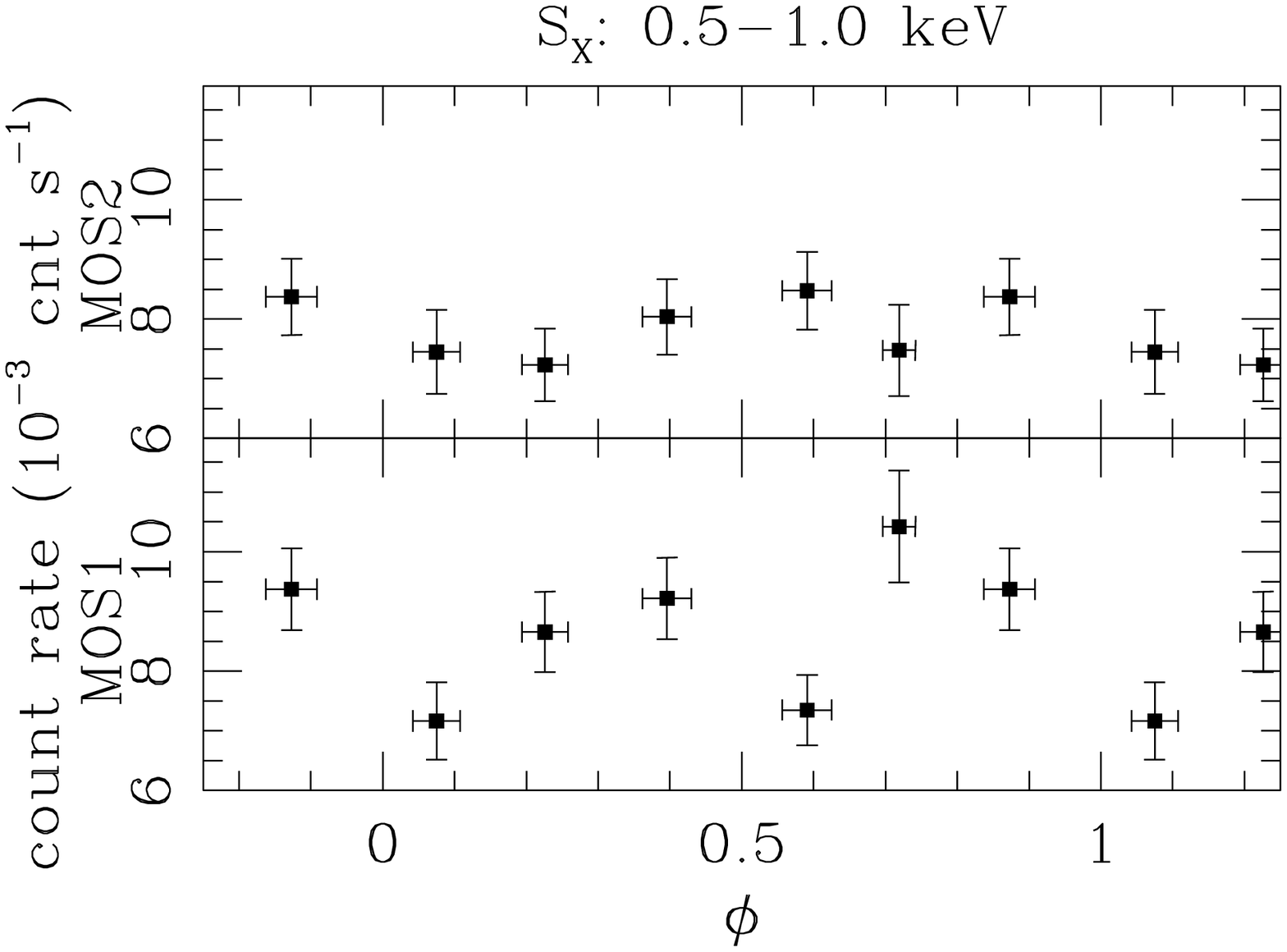}
\includegraphics[width=.4\textwidth]{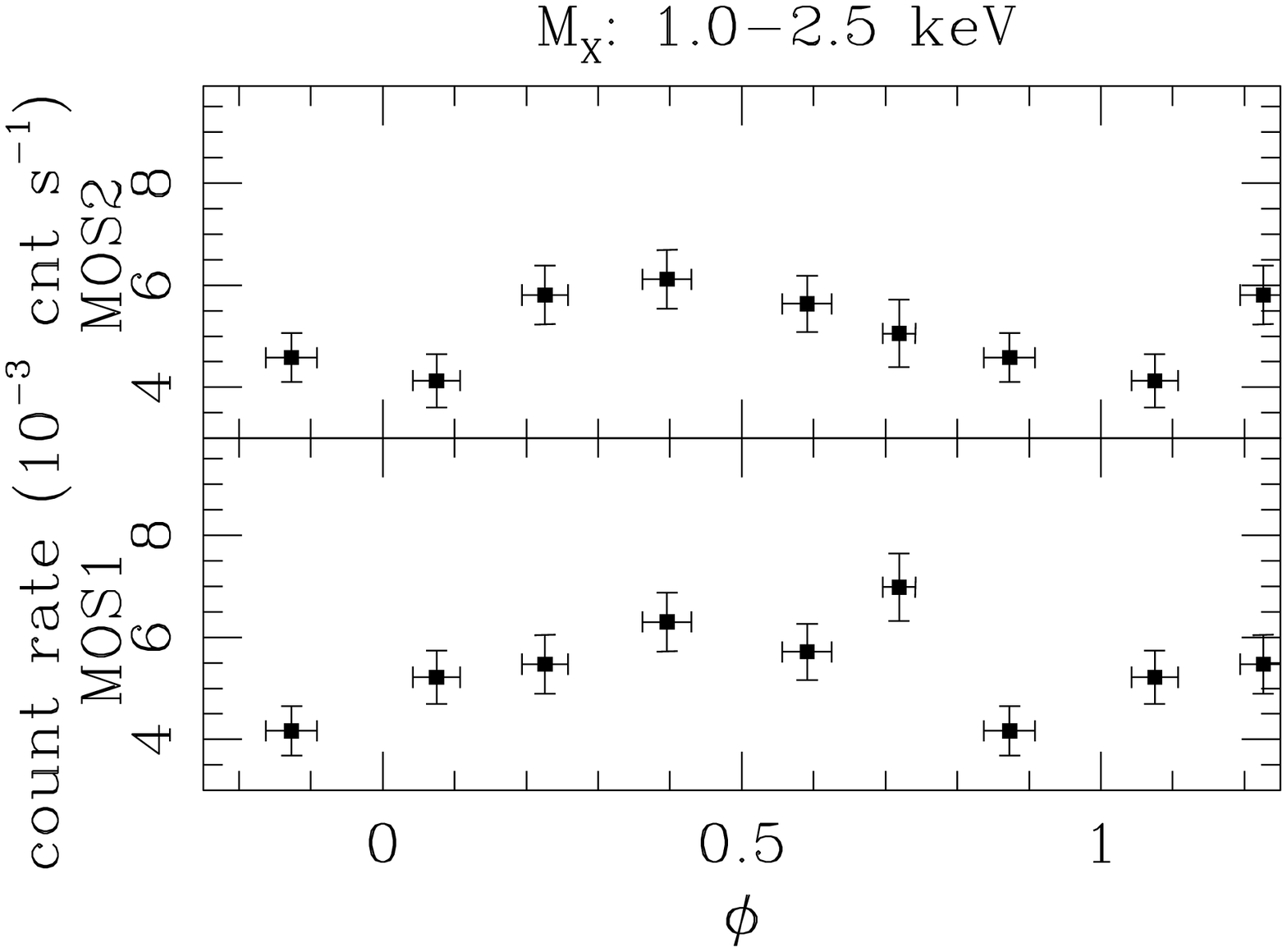}\hspace*{5mm}
\includegraphics[width=.4\textwidth]{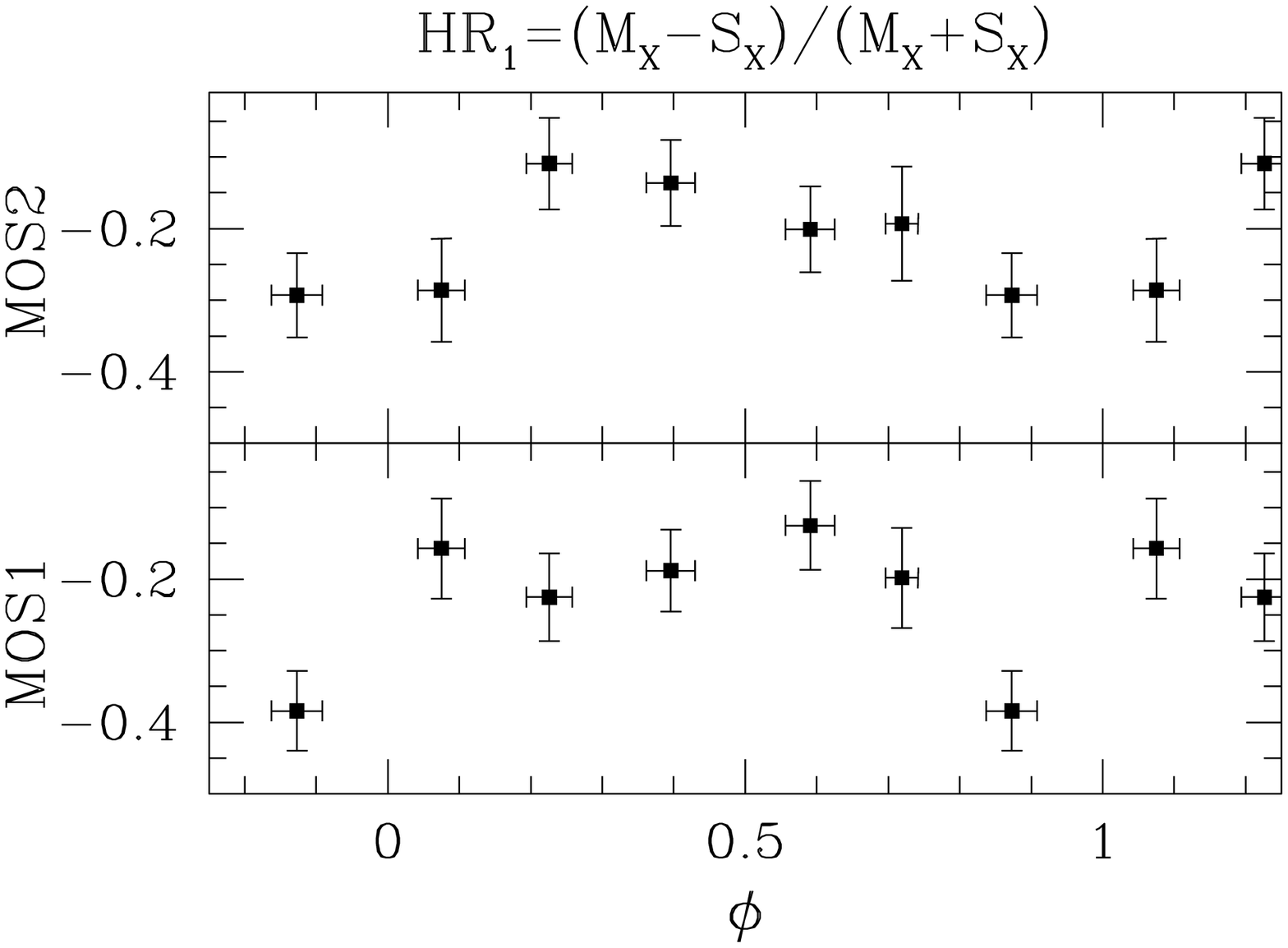}
\caption{Broad band X-ray light curves in the two \epicmos\ instruments. The different energy ranges considered are indicated on top of each panel. The vertical bars give the $\pm$1-$\sigma$ uncertainty on the background-corrected count rates. The horizontal bars indicate the duration of each pointing. The bottom right panel displays the evolution of the source  hardness ratio with the phase, as defined on top of the panel.}
\label{fig: xlc}
\end{figure*}

\subsection{Masses and orbital inclination}

Fig. \ref{fig: hr} presents the location of the stars in the H-R diagram. The evolutionary age of both components is about 4 to 5~Myr, which is in agreement with previous cluster age determinations from photometric studies  \citep{PHYB90, BL95, RCB97, BVF99}. From a rough interpolation of the tracks, the main sequence primary had an initial mass of about 25~\msol, corresponding to a present  mass close to 24.5~\msol, slightly larger than typical O9V masses (which suggests again that the primary component has reached an intermediate stage of evolution between classes V and III). Similarly, the secondary initial mass was about 18.5~\msol. Its present mass should be about 18.2~\msol.

Comparing the obtained minimal mass of Table \ref{tab: orbit} with typical O9V  masses suggests an orbital inclination $i$ of about 70\degr. Comparing with O9III masses would rather yield $i\sim60$\degr. Using the constraints deduced on the components radii, we found that the limiting inclination for which eclipses may or not occur in the system lies around 55-65\degr. Clearly, \astrobj{HD 152218} is a limiting case. Photometric monitoring is required to search for the signature of eclipses or ellipsoidal deformation that could help to constrain the inclination of the system. Finally, we used the formula of \citet{Egg83} to estimate the Roche lobe radii. We found that, according to the previous results, both stars should presently fill about 17\% of their Roche lobe volume.

\begin{sidewaystable*}
\begin{center}
\tiny
\caption{ Best-fitting models and resulting X-ray fluxes at Earth. The 1-T model used was {\tt wabs$_\mathrm{ISM}$ * wabs($N_\mathrm{H}$) * mekal(k$T$)} while the 2-T model corresponds to  {\tt wabs$_\mathrm{ISM}$ * (mekal(k$T_1$) + wabs($N_\mathrm{H, 2}$) * mekal(k$T_2$))}. The term {\tt wabs$_\mathrm{ISM}$} was fixed to the interstellar value ($N_\mathrm{H, ISM}=0.29\times 10^{22}$\,cm$^{-2}$). Columns 1 and 2 give the phase and the observation number. Columns 3 to 5 provide the best-fit parameters:  $N_\mathrm{H}$, the absorbing column;  $kT$, the model temperature; $norm$, the normalisation factor ($norm =\frac{10^{-14}}{4\pi d^2}\int n_\mathrm{e} n_\mathrm{H} dV$ with $d$, the distance to the source -- in cm --, $n_\mathrm{e}$ and $n_\mathrm{H}$, the electron and hydrogen number densities -- in cm$^{-3}$). The quoted upper and lower values provide the limits of the 90\% confidence interval. Column 6 lists the corresponding reduced chi-square and the associated number of degrees of freedom (d.o.f). Columns 7 to 10 provide the observed fluxes in the 0.5 - 10.0\,keV energy band and in the S$_\mathrm{X}$ (0.5 - 1.0\,keV), M$_\mathrm{X}$ (1.0 - 2.5\,keV) and H$_\mathrm{X}$ (2.5 - 10.0\,keV) bands respectively while Columns 11 to 14 provide the same fluxes, but corrected for the adopted ISM absorbing column. Column 15 finally provides the X-ray luminosities in the 0.5 - 10.0 keV band, assuming $DM=11.07$.
} 
\label{tab: Xspec}
\begin{tabular}{c c c c c c c c c c c c c c c}
\hline
$\phi_\mathrm{HeI}$ & Obs. \# & $N_\mathrm{H}$ & k$T$ & $norm$ & $\chi^2_{\nu}$ (d.o.f.) & $f_\mathrm{X}$ & $f_\mathrm{X,S}$ & $f_\mathrm{X,M}$  & $f_\mathrm{X,H}$  & $f_\mathrm{X}^\mathrm{corr.}$ & $f_\mathrm{X,S}^\mathrm{corr.}$ & $f_\mathrm{X,M}^\mathrm{corr.}$  & $f_\mathrm{X,H}^\mathrm{corr.}$ & $\log L_\mathrm{X}$ \\
                   &         & ($10^{22}$\,cm$^{-2}$) & (keV) & (cm$^{-5}$) &           & \multicolumn{4}{c} ($10^{-14}$\,erg\,cm$^{-2}$\,s$^{-1}$)               & \multicolumn{4}{c} ($10^{-14}$\,erg\,cm$^{-2}$\,s$^{-1}$)                                                                      & (\ergs)           \\
$[1]$              & $[2]$   & $[3]$         & $[4]$& $[5]$  & $[6]$                  & $[7]$         & $[8]$           & $[9]$            & $[10]$           & $[11]$                      & $[12]$                        & $[13]$                         & $[14]$                        & $[15]$             \\
\hline
\vspace*{-3mm}\\
\multicolumn{15}{c}{Single-temperature (1-T) model}\\
\hline
\vspace*{-3mm}\\
  0.591  & 1 & 0.10$_{0.02}^{0.19}$ & 0.53$_{0.47}^{0.58}$ & 1.16$_{0.84}^{1.56} \times 10^{-4}$ & 1.17 (47) & 8.54 & 4.72 & 3.68 & 0.13  \vspace*{1mm} & 20.61 & 14.78 & 5.69 & 0.14 &  31.82 \\
  0.719  & 2 & 0.11$_{0.00}^{0.22}$ & 0.53$_{0.46}^{0.60}$ & 1.27$_{0.85}^{1.84} \times 10^{-4}$ & 0.82 (32) & 9.27 & 5.07 & 4.05 & 0.15  \vspace*{1mm} & 22.20 & 15.80 & 6.24 & 0.16 &  31.85 \\
  0.873  & 3 & 0.00$_{0.00}^{0.03}$ & 0.61$_{0.57}^{0.64}$ & 7.63$_{7.04}^{8.51} \times 10^{-5}$ & 1.46 (47) & 8.11 & 4.70 & 3.25 & 0.15  \vspace*{1mm} & 19.80 & 14.63 & 5.00 & 0.16 &  31.80 \\
  0.075  & 4 & 0.38$_{0.25}^{0.53}$ & 0.24$_{0.19}^{0.31}$ & 8.40$_{0.00}^{29.8} \times 10^{-4}$ & 0.63 (35) & 7.86 & 4.87 & 2.98 & 0.01  \vspace*{1mm} & 22.24 & 17.47 & 4.75 & 0.01 &  31.85 \\
  0.226  & 5 & 0.40$_{0.26}^{0.50}$ & 0.25$_{0.21}^{0.32}$ & 9.20$_{0.00}^{23.0} \times 10^{-4}$ & 1.35 (44) & 8.36 & 5.01 & 3.33 & 0.01  \vspace*{1mm} & 22.90 & 17.58 & 5.31 & 0.01 &  31.87 \\
  0.396  & 6 & 0.38$_{0.26}^{0.47}$ & 0.24$_{0.22}^{0.32}$ & 9.23$_{0.00}^{17.3} \times 10^{-4}$ & 1.48 (50) & 8.70 & 5.36 & 3.32 & 0.01  \vspace*{1mm} & 24.47 & 19.15 & 5.31 & 0.01 &  31.89 \\
\hline                                                                                                                                                    	     			            
\vspace*{-3mm}\\                                                                                                                                         
  -- & Merged & 0.01$_{0.00}^{0.05}$ & 0.58$_{0.57}^{0.60}$ & 8.62$_{8.06}^{9.91} \times 10^{-5}$ & 1.38 (169)& 8.75 & 5.11 & 3.49 & 0.15 \vspace*{1mm} & 21.55 & 16.01 & 5.39 & 0.15 &  31.84 \\
\hline
\end{tabular}
\end{center}
\end{sidewaystable*}

\begin{sidewaystable*}
\begin{center}
\tiny
Table 6. {\it Continue}\\
%\label{tab: Xspec}
\begin{tabular}{c c c c c c c c c c c c c c c}
\hline
$\phi_\mathrm{HeI}$ & Obs. \# & $N_\mathrm{H}$ & k$T$ & $norm$ & $\chi^2_{\nu}$ (d.o.f.) & $f_\mathrm{X}$ & $f_\mathrm{X,S}$ & $f_\mathrm{X,M}$  & $f_\mathrm{X,H}$  & $f_\mathrm{X}^\mathrm{corr.}$ & $f_\mathrm{X,S}^\mathrm{corr.}$ & $f_\mathrm{X,M}^\mathrm{corr.}$  & $f_\mathrm{X,H}^\mathrm{corr.}$ & $\log L_\mathrm{X}$ \\
                   &         & ($10^{22}$\,cm$^{-2}$) & (keV) & (cm$^{-5}$) &           & \multicolumn{4}{c} ($10^{-14}$\,erg\,cm$^{-2}$\,s$^{-1}$)               & \multicolumn{4}{c} ($10^{-14}$\,erg\,cm$^{-2}$\,s$^{-1}$)                                                                      & (\ergs)           \\
$[1]$              & $[2]$   & $[3]$         & $[4]$& $[5]$  & $[6]$                  & $[7]$         & $[8]$           & $[9]$            & $[10]$           & $[11]$                      & $[12]$                        & $[13]$                         & $[14]$                        & $[15]$             \\
\hline
\vspace*{-3mm}\\
\multicolumn{15}{c}{Two-temperature (2-T) model}\\
\hline
\vspace*{-3mm}\\
 0.591 & 1 & --                   & $0.34^{0.54}_{0.24}$ & $9.06^{11.3}_{5.59} \times 10^{-5}$ \vspace*{1mm}\\
       &   & $0.57^{1.13}_{0.24}$ & $0.68^{0.87}_{0.53}$ & $9.97^{16.0}_{5.52} \times 10^{-5}$ & 0.95 (45) &  9.31 & 5.08 & 3.96 & 0.27 \vspace*{1mm}& 24.64 & 18.42 &  5.93  & 0.29 & 31.90 \\
 0.873 & 3 & --                   & $0.24^{0.29}_{0.14}$ & $1.11^{2.69}_{0.68} \times 10^{-4}$ \vspace*{1mm}\\
       &   & $0.10^{0.36}_{0.00}$ & $0.71^{0.83}_{0.61}$ & $5.88^{8.61}_{3.88} \times 10^{-5}$ & 1.07 (45) &  9.13 & 5.75 & 3.18 & 0.19 \vspace*{1mm}& 28.67 & 23.65 &  4.82  & 0.20 & 31.96 \\ 
 0.226 & 5 & --                   & $0.32^{0.41}_{0.24}$ & $8.98^{11.7}_{5.29} \times 10^{-5}$ \vspace*{1mm}\\
       &   & $0.36^{0.68}_{0.11}$ & $0.71^{0.99}_{0.57}$ & $7.24^{12.0}_{4.22} \times 10^{-5}$ & 1.19 (42) &  9.12 & 5.22 & 3.67 & 0.24 \vspace*{1mm}& 24.84 & 19.05 &  5.54  & 0.25 & 31.90 \\ 
 0.396 & 6 & --                   & $0.31^{0.38}_{0.26}$ & $1.12^{1.36}_{0.88} \times 10^{-4}$ \vspace*{1mm}\\
       &   & $0.63^{1.00}_{0.37}$ & $0.71^{0.83}_{0.57}$ & $1.06^{1.65}_{0.73} \times 10^{-4}$ & 0.97 (48) &  9.99 & 5.58 & 4.08 & 0.32 \vspace*{1mm}& 27.70 & 21.29 &  6.08  & 0.34 & 31.95 \\ 
\hline
\vspace*{-3mm}\\	       
  -- & Merged & --                   & $0.31^{0.34}_{0.29}$ & $9.78^{10.8}_{8.74} \times 10^{-5}$ \vspace*{1mm}\\
       &   & $0.39^{0.50}_{0.28}$ & $0.71^{0.76}_{0.65}$ & $7.83^{9.20}_{6.56} \times 10^{-5}$ & 0.89 (164)&  9.55 & 5.47 & 3.82 & 0.25 \vspace*{1mm}& 26.26 & 20.23 &  5.77  & 0.26 & 31.93 \\
\hline
\end{tabular}
\end{center}
\end{sidewaystable*}

%_____________________ XMM OBSERVATIONS ____________________________

\section{\xmm\ observations} \label{sect: xmm}

\subsection{Light curves and spectra}

With a typical count rate of about $12\times10^{-3}$~\cnts\ in  the two \mos\ instruments, \astrobj{HD 152218} is a relatively bright X-ray source. Because the count rates in the hard energy band (2.5-10.0keV) are very low, thus associated with large uncertainties, we restrain our analysis to energies below 2.5~keV.  Fig. \ref{fig: xlc} presents the obtained broad band light curves in the total band and in the S$_\mathrm{X}$ [0.5-1.0~keV]  and M$_\mathrm{X}$ [1.0-2.5~keV] bands. Except for Obs. 2 at $\phi\approx0.7$, the two instruments give very consistent results, indicating a slight modulation with an increase of the X-ray flux of about 30\% at apastron compared to periastron. 

\begin{figure}
\centering
\includegraphics[angle=-90,width=\columnwidth]{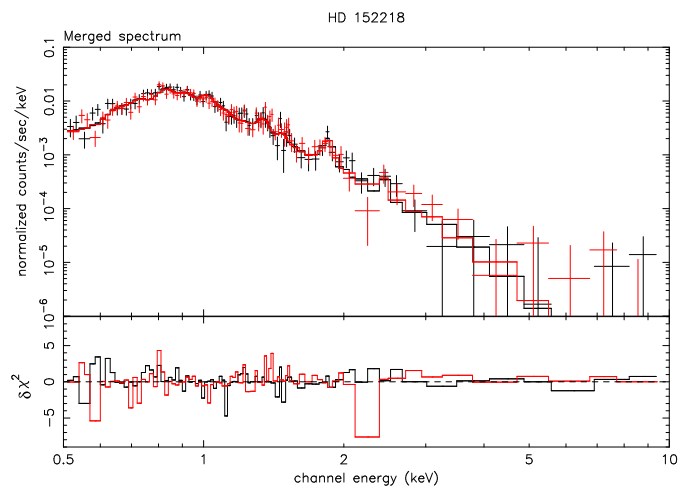}
\caption{{\bf Top panel:} Merged \epicmos1 (black) and merged \epicmos2 (light grey -- hardcopy edition -- or red -- electronic edition) X-ray spectra of \astrobj{HD 152218} and best-fit 2T models. {\bf Lower panel:} Individual contributions of the different energy bins to the $\chi^2$. The contributions are carried over with the sign of the deviation (in the sense data minus model).}
\label{fig: 2T}
\end{figure}

To constrain the physical properties of the emitting plasma, we adjusted a series of optically thin thermal plasma (\mek)  models using the \xspec\ software \citep{Arn96}.  We simultaneously fitted the two \mos\ spectra of each pointing, adopting an interstellar column of absorbing matter (neutral hydrogen) of $N_\mathrm{H, ISM}=0.29\times10^{22}$~cm$^{-2}$. This value was obtained using $E(B-V)=0.50$ and the formula of \citet{BSD78}. Except for Obs 2 and 4, for which the effective exposure time was reduced by high background events, better results are obtained using two-temperature (2-T) \mek\ models.  
 Table \ref{tab: Xspec} summarizes the best-fit values and gives the intrinsic fluxes and luminosities. There is no obvious modulation of the spectral parameters. The modulation of the hardness ratios (Fig. \ref{fig: xlc}) however suggests that the observed emission is slightly harder when the flux level is high, and lower otherwise.    The merged spectra are well described by 2-T models with k$T_1=0.31$~keV and k$T_2=0.71$~keV. Fig. \ref{fig: 2T} displays these spectra together with the best-fit 2-T models. It reveals that a hard emission component ($>5$~keV) might be present, but clearly the SNR is far insufficient to reliably constrain its properties.

\subsection{\astrobj{HD 152218} wind properties}
We used the constraints on the physical properties deduced in Sect. \ref{sect: physic} together with the mass loss recipes of \citet{VdKL00, VdKL01} to estimate the properties of the winds in \astrobj{HD 152218}.  We obtained $v_{\infty, 1}=2280$~\kms, $v_{\infty, 2}=2340$~\kms, $\log \dot{M}_1=-6.59$ (\msolyr) and $\log \dot{M}_2=-7.37$ (\msolyr). The estimated terminal velocities are only slightly larger than the 2050~\kms\ value measured by \citet{HSHP97}.
Using these parameters with a $\beta=1$ acceleration law for the winds, we could not establish a ram pressure equilibrium at all phases, indicating that the wind collision structure could be quite unstable. We suspect that the wind velocity law might be strongly affected by radiative inhibition \citep{StP94}. The latter  may even govern the flux of colliding matter. Clearly, only a detailed modeling could allow to get an insight into the wind collision in this system. This is far beyond the goal of the present paper. We will thus limit ourselves to the observational result that the flux is slightly variable and that the maximum of emission is observed near apastron while the minimum is reached at periastron. As \astrobj{HD 152218} is a close binary, the wind interaction will inevitably take place within the acceleration region of the winds. We thus expect the collision  to be stronger at apastron, because of the higher pre-shock velocity reached by the wind at this particular phase. However, we caution that more subtle aspects (such as radiative inhibition) probably affect the strength of the winds, and might thus govern the actual position and shape of the  on-going wind interaction.

%___________________________ CONCLUSIONS ______________________________

\section{Summary} \label{sect: ccl}

We have presented the results of a long-term spectroscopic monitoring campaign on the massive binary \astrobj{HD 152218}. The last two investigations on the object were presenting conflicting results. The present data  complete  the existing observational sets and allow to definitely overcome the ambiguity on the orbital period of the system, bringing to firm ground the period of  5.604~d. From our analysis, we reject the shorter period and higher eccentricity proposed by \citet{GM01} and rather confirm the earlier work of \citet{SLP97}, based on IUE observations.  Our newly derived orbital solution is characterized by a slightly lower eccentricity than previously accepted: $e=0.259\pm0.006$. The system is most probably formed by an O9 sub-giant and an O9.7 main sequence star. We derived minimal masses of $15.82 \pm 0.26$~\msol\ and   $12.00 \pm 0.19$~\msol\ and we constrained the component radii to values of $R_1=10.1\pm1.0$~\rsol\ and $R_2=8.9\pm2.0$~\rsol.  These values are in good agreement with previous findings and further indicate that \astrobj{HD 152218} should have an orbital inclination of about 60--70\degr. This corresponds to a limiting case in which eclipses or ellipsoidal variations might be observed. 

We also report the results of the monitoring of the system in the X-rays thanks to \xmm\ observations of the cluster. The averaged X-ray spectrum is relatively soft. It is well reproduced by a 2-T \mek\ model with component temperatures about 0.3 and 0.7~keV. We showed that the system presents an increase of its  X-ray flux  of about 30\% near apastron compared to periastron. We note that this could be the signature of an ongoing wind-wind interaction process occurring within the wind acceleration region. Such a scenario indeed predicts a stronger shock, and thus a larger X-ray emission from the colliding zone, when the separation between the components is larger, allowing the winds to reach larger pre-shock velocities. This is also supported by the apparent modulation of the hardness ratio, which seems higher when the emission level is stronger. Preliminary computations however indicate that no simple modeling, reproducing the main properties of the interaction, can be obtained. By contrast, the same computations rather point towards second order effects, such as radiative inhibition, to play a major role in this system. These effects could be strong enough to govern the structure and shape of the interaction. HD~152218 might thus be an excellent case to test and/or refine wind collision models, including detailed physics on the wind acceleration and inhibition.

%_____________ ACKNOWLEDGMENTS __________________________________________

\section*{Acknowledgments}

 The Li\`ege team is greatly indebted towards the `Fonds de la Recherche Scientifique` (FNRS), Belgium, for multiple supports. Part of this work was also supported by the PRODEX XMM-OM and Integral Projects, as well as contracts P4/05 and P5/36 `P\^ole d'Attraction Interuniversitaire' (Belgium).

% \appendix

% \section{}
% \label{}

% Bibliographic references with the natbib package:
% Parenthetical: \citep{Bai92} produces (Bailyn 1992).
% Textual: \citet{Bai95} produces Bailyn et al. (1995).
% An affix and part of a reference:
%   \citep[e.g.][Ch. 2]{Bar76}
%   produces (e.g. Barnes et al. 1976, Ch. 2).

%\bibliographystyle{elsart-harv.bst}
%\bibliography{/home/hsana/disk-externe/LIEGE_PAPERS/XMM_CAT_PAPER4pdf/ngc6231_Xcat}

%\begin{thebibliography}{}

% \bibitem[Names(Year)]{label} or \bibitem[Names(Year)Long names]{label}.
% (\harvarditem{Name}{Year}{label} is also supported.)

%\bibitem[]{}

%\end{thebibliography}

\end{document}